\documentclass[12pt]{article}
\usepackage{amsmath}
\usepackage{amssymb}
\usepackage{amsthm}
\usepackage{bbm,mathtools}
\usepackage{theoremref}
\usepackage{enumitem}
\usepackage{kpfonts}
\usepackage{xparse}
\usepackage{accents}
\usepackage{mathrsfs}
\usepackage{calrsfs}
\usepackage{subcaption}
\usepackage{float}
\usepackage[colorlinks=true,linkcolor=blue,citecolor=blue,urlcolor=blue]{hyperref}

\newtheorem{theorem}{Theorem}
\newtheorem{lemma}[theorem]{Lemma}
\newtheorem{corollary}[theorem]{Corollary}
\newtheorem{proposition}[theorem]{Proposition}

\newtheorem{remark}{Remark}
\theoremstyle{definition}
\newtheorem{definition}{Definition}
\newtheorem{assumption}{Assumption}

\def\rig#1{\mathcal{R}(#1)}
\def\cut#1{\mathcal{L}^n(#1)}
\def\cutabs#1{\mathcal{L}(#1)}

\newcommand{\rn}{\mathbb{R}} 
\newcommand{\nn}{\mathbb{N}} 

\newcommand{\vol}{\mathsf{v}} 
\newcommand{\trig}{\mathcal{T}} 
\newcommand{\trigb}{\mathcal{T}^{\circledast}} 
\newcommand{\fourier}{\mathcal{F}} 
\newcommand{\fourierb}{\mathcal{F}^{\circledast}} 
\newcommand{\fourierj}{\dddot{\mathcal{F}}} 
\newcommand{\fejer}{{\mathbf{F}}} 

\NewDocumentCommand{\resi}{o}{\mathsf{R}_{\IfValueTF{#1}{#1}{\pi}}(q,N)}
\NewDocumentCommand{\residis}{o}{\mathsf{R}_{\IfValueTF{#1}{#1}{\pi}}(q,n,N)}

\newcommand{\price}{\mathsf{P}} 
\newcommand{\pow}{\mathsf{g}}  
\newcommand{\bohr}[2]{#1 \mathbin{\circledast} #2} 
\newcommand{\aux}{\mathsf{T}} 

\newcommand{\integrabilityratea}{\mathsf{h}}  
\newcommand{\integrabilityratejumps}{\mathsf{j}}  
\newcommand{\integrabilityratejumpsa}{\mathsf{q}}  
\newcommand{\poissonp}{\mathsf{N}}  

\NewDocumentCommand{\bohrp}{ m m o }{%
	#1 \mathop{\circledast}\limits_{\IfValueTF{#3}{#3}{N}} #2%
}

\NewDocumentCommand{\triplenorm}{O{p} m}{%
\left\lvert\!\left\lvert\!\left\lvert #2
	\right\rvert\!\right\rvert\!\right\rvert%
	\IfValueT{#1}{_{#1}}%
}

\begin{document}
\title{Fourier Estimation of spot volatility: Unbounded coefficients and jumps in the price process}
\author{L.J. Espinosa González\thanks{Instituto de Matem\'aticas, Unidad  Cuernavaca, Universidad Nacional Aut\'onoma de M\'exico, M\'exico}
\and
Erick Trevi\~no-Aguilar\thanks{Instituto de Matem\'aticas, Unidad  Cuernavaca, Universidad Nacional Aut\'onoma de M\'exico, M\'exico.}}

\maketitle

\begin{abstract}
We study a Fourier-based estimator of the spot volatility of a semimartingale, originally introduced by Malliavin and Mancino, and extend its asymptotic theory. First, we remove the uniformly bounded volatility assumption and establish almost sure convergence of the estimated Fourier coefficients under mild integrability conditions. We further derive explicit convergence rates that yield almost sure convergence of the reconstructed spot volatility via Fejér inversion.

We then extend the analysis to processes with jumps. In this setting, we provide a precise decomposition of the Fourier coefficient estimation error into diffusion–diffusion, diffusion–jump, jump–diffusion, and jump–jump interaction terms, and obtain $L^p$-rates for each component. We show that the estimated coefficients converge almost surely to the sum of the true volatility coefficients and an explicit jump-induced term. Moreover, we prove that an appropriate rescaling of the trigonometric reconstruction yields almost sure pathwise recovery of the quadratic jump process.
\end{abstract}

\maketitle

\section{Introduction}
In this paper we study a Fourier-based estimator of the spot volatility, originally introduced by Malliavin and Mancino, and extend its asymptotic theory. Our main objective is to establish strong pathwise reconstruction results for the spot volatility  using Fourier-based methods. We prove convergence of trigonometric polynomials to the volatility path under mild integrability conditions, allowing the volatility process to be unbounded. We further derive explicit convergence rates for the estimation error of Fourier coefficients and of the reconstructed paths, with speeds sufficient to obtain almost sure convergence rather than merely convergence in probability. To the best of our knowledge, this is new even in the setting of continuous semimartingales.

We then extend the analysis to the case of c\`adl\`ag processes with jumps. In this setting, we show that the effect of jumps on the Fourier estimator can be characterized precisely under additional mild integrability conditions. Using a martingale decomposition and Burkholder--Davis--Gundy type inequalities, we derive $L^p$-rates for the interaction terms between the continuous and jump components of the underlying process. We prove that a minor correction of rescaling type allows one to asymptotically recover the quadratic jump process from the Fourier reconstruction. This result provides an explicit description of the jump-induced bias in the estimated Fourier coefficients and yields an almost sure pathwise recovery theorem for jump variation.

The effect of jumps in Fourier-based volatility estimators is a question of both theoretical and practical relevance. It has been emphasized in several works that jumps play a significant role in high-frequency data; see, for instance, \cite{Carr2002}. The problem of understanding how jumps affect the Malliavin--Mancino estimator has been explicitly raised by previous authors; see \cite[Remark~3.2]{CUCHIERO2015} and \cite[p.~369]{Jacod2019b}. The present paper provides an analysis of this question while retaining the original Bohr convolution structure of the estimator.

We focus on the one-dimensional case, both in the number of processes and in the number of stochastic sources, mainly for notational simplicity and because the fundamental technical difficulties already appear in this setting.

The paper is organized as follows. In Section~\ref{labsec:priceprocess} we fix a dynamic for an underlying process with continuous paths. In Section~\ref{labsec:Preliminaries} we present general preliminaries. In Section~\ref{labsec:fouriercoeff} we introduce the Fourier estimator due to Malliavin and Mancino \cite{Malliavin2002, Malliavin2009}.

In Section~\ref{labsec:continuousobservation} we consider a continuous-time dynamic and prove uniform convergence of Fourier--Fej\'er trigonometric polynomials constructed from two different systems of coefficients. The first system corresponds to the idealized case in which the  spot volatility is observed and serves as a theoretical benchmark. The second system is defined via the Bohr convolution of coefficients, which is the core of the Malliavin--Mancino estimator and is constructed from only observable quantities.

Assuming continuous observation of an underlying path, we then extend the analysis in Section~\ref{labsec:discreteobservation} to the more realistic setting of discrete observation again with access to the quadratic variation density path. In Section~\ref{labsec:continuousobservationJumps} we further extend the continuous-observation results to a dynamic with jumps in the underlying process. We prove that, in this setting, rescaled trigonometric polynomials obtained through Bohr convolution converge almost surely to the quadratic jump process.

Finally, in Section~\ref{labsubs:cppnumericalillust} we provide numerical illustrations of the theoretical results.

\section{The model}\label{labsec:priceprocess}
Since our approach is based on Fourier series, we take the time index to be the compact interval $[-\pi,\pi]$. Let $(\Omega,\mathcal{F},\mathbb{F}=\{\mathcal{F}_t\}_{-\pi \le t \le \pi},\mathbb{P})$ be a filtered probability space satisfying the usual conditions and supporting a standard Brownian motion $W$.

We consider a continuous semimartingale $H=\{H_t\}_{-\pi \le t \le \pi}$ defined as the unique strong solution to the stochastic differential equation
\begin{align}
	dH_t &= \sigma_t \, dW_t, \qquad H_{-\pi} = x \in \mathbb{R}, \label{labEqdyn}
\end{align}
where $\sigma=\{\sigma_t\}_{-\pi \le t \le \pi}$ is a progressively measurable process with continuous sample paths. In line with \cite{Malliavin2002,Malliavin2009}, $H$ may be interpreted as a log-price process.

\begin{remark}
	For notational simplicity, we do not include a drift term in \eqref{labEqdyn}. As shown in \cite{Malliavin2009}, any drift component does not contribute to the Fourier estimator and therefore plays no role in the asymptotic analysis developed below.
\end{remark}

We impose the following integrability condition on the diffusion coefficient.

\begin{assumption}\thlabel{lab:integrabilityforsigma}
	There exists $\integrabilityratea>0$ such that
	\[
	\mathbb{E}\!\left[ \int_{-\pi}^{\pi} |\sigma_z|^{\integrabilityratea} \, dz \right] < \infty.
	\]
\end{assumption}

\begin{remark}\thlabel{labrem:easydiffusions}
More specific information on the exponent $\integrabilityratea$ in \thref{lab:integrabilityforsigma} will be given in the convergence results below; see \thref{labthm:coeffConvergence} and \thref{labthm:PathUC}.
	
If the diffusion coefficient is of the form $\sigma_t = f(t,H_t)$ for a measurable function $f$, then the stochastic differential equation
\[
dH_t = f(t,H_t) \, dW_t
\]
admits a unique strong solution under mild regularity conditions on $f$; see \thref{labtheasydiffusions} in the appendix. In this case, $\sigma$ satisfies \thref{lab:integrabilityforsigma} for arbitrarily large exponents $\integrabilityratea$; see \thref{labtheasydiffusionscorollaryintegrability}.
\end{remark}

We define the spot volatility  $\vol = \{\vol_t\}_{-\pi \le t \le \pi}$ by
\begin{align}\label{labEq:martingalepartofp}
	\vol_t := \sigma_t^2.
\end{align}

\section{Preliminaries}\label{labsec:Preliminaries}
We start with some notation.  Let $a:=\{a_n\}_{n \in \nn}$ and  $b:=\{b_n\}_{n \in \nn}$ be two sequences of non negative real numbers. We write $a=O(b)$ if there exists a constant $K>0$ such that $a_n \leq K b_n$ for $n \in \nn$. We denote by $\imath$ the imaginary number, solution of $x^2=-1$.\\

We denote by $D_N:[-\pi,\pi] \to \mathbb{C}$ the \textbf{Dirichlet kernel} that includes $2N + 1$ harmonics. It is defined by 
\begin{equation}\label{labEq:DirichletKernel}
D_N(t):= \sum_{|l| \leq N} \exp (\imath l t).
\end{equation}
We also introduce the \textbf{rescaled Dirichlet kernel}:
\begin{equation}\label{labEq:RescaledDirichletKernel}
\tilde D_N(t):=\frac{1}{2N +1} D_N(t).
\end{equation}

We denote by $S_N[\phi]$ the {partial sum} of the Fourier series of a function $\phi$:
\[
S_N[\phi](t):=\sum_{|n| \leq N} \fourier[\phi](n) e^{\imath n t }.
\]
The Dirichlet kernel allows us to express the partial sums of  Fourier series  as a convolution:
\[
S_N[\phi](t)=\phi * D_N(t):=\frac{1}{2\pi} \int_{-\pi}^{\pi}\phi(s)D_N(t-s)ds;
\]
see \cite[p. 44]{SteinShakarchi2003}.\\

The Fej\'er kernel is defined by 
\[
\fejer_N(t):=\frac{1}{N}\sum_{j=0}^{N-1} D_N(t).
\]

Now we continue with a few basic concepts for martingales. For a process $X$ we denote, as usual, its running supremum by $X^*$, hence 
\begin{equation}\label{labeq:runningsupremum}
X^*_t:= \sup_{-\pi \leq s \leq t}|X_s|.
\end{equation}
For $K$ a local martingale with cadlag paths, and $K_{-\pi}=0$ we denote by $\left[K\right] $ its quadratic variation. Recall the Burkholder-Davis-Gundy inequality (BDG inequality) in the following form. For $p>1$ there exist positive constants $c_p$ and $C_p$ such that for a stopping time $\tau$
\[
c_p E[\left[K \right]_{\tau}^{p/2}] \leq E[(K^*_{\tau})^p] \leq C_p E[\left[K \right]_{\tau}^{p/2}];
\]
see \cite[Theorem 10.36]{hwy92}.

If $K$ has continuous paths, we consider the BDG inequality in the following form. Denote by $\left\langle K \right\rangle$ the predictable quadratic variation of $K$. For $p>0$ there exist positive constants $c_p$ and $C_p$ such that for $\tau$ a stopping time
\[
c_p E[\left\langle K \right\rangle_{\tau}^{p/2}] \leq E[(K^*_{\tau})^p] \leq C_p E[\left\langle K \right\rangle_{\tau}^{p/2}];
\]
see \cite[(4.1)]{Revuz2005}.

We will also work with complex-valued processes. For $X=\Re X + \imath \Im X$ a complex valued process, we adapt  the notation for the running supremum \eqref{labeq:runningsupremum} as follows:
\begin{align}
X^*_t&:=\sup_{-\pi \leq s \leq t} |\Re X_s| + \sup_{-\pi \leq s \leq t} |\Im X_s|.\label{labeq:runningsupremumcomplexc}
\end{align}
Furthermore, for $p \geq 2$, we denote by $\triplenorm{X}$ the following norm
\begin{equation}\label{labeq:triplenorm}
\triplenorm{X}:=\sqrt[p]{E\left[\left( \sup_{-\pi\leq s \leq \pi} |\Re X_s|\right) ^p\right]}+\sqrt[p]{E\left[\left( \sup_{-\pi \leq s \leq \pi} |\Im X_s|\right) ^p\right]}.
\end{equation}
Note the triangle inequality in this notation:
\[
\left\|X^*_t\right\|_{L_p} \leq \triplenorm{X}.
\]
\section{The Fourier coefficients of volatility}\label{labsec:fouriercoeff}
\subsection{Basic definitions on Fourier analysis}
For $q \in \mathbb{Z}$, the $q$-th \textbf{Fourier coefficient}  of  a function $\phi:[-\pi, \pi] \to \rn$ is denoted $\mathscr F[\phi](q)$. It is defined by
\[
\fourier [\phi](q):=\frac{1}{2 \pi} \int_{-\pi}^{\pi} e^{- \imath q t}\phi(t)dt.
\]
The \textbf{Fourier coefficient of a differential} is given by 
\[
\fourier [d\phi](q):=\frac{1}{2 \pi}  \int_{-\pi}^{\pi} e^{- \imath q t} d\phi_t, \mbox{ for } q \in \mathbb Z.
\]
Clearly, further conditions on $\phi$ are necessary for this definition to be correct. For  Riemann-Stieltjes integration under a pathwise perspective it is sufficient that $\phi$ is cadlag since the exponential is a smooth function. Indeed this follows by integration by parts; see \cite[Theorem 7.6]{Apostol1974}. For the stochastic counterpart see e.g., \cite[Proposition (2.13)]{Revuz2005} and \cite[Theorem II.17]{Protter2005}.
\subsection{The Bohr convolution}
Malliavin and Mancino \cite{Malliavin2009} define the concept of \textbf{Bohr convolution}.  We recall it now.  Let $u,v :\mathbb Z \to \mathbb{C}$ be two complex-valued functions.   Bohr's convolution of $u$ and $v$ denoted  $\bohr{u}{v}$ is defined if the following limit exists for each $q \in \mathbb Z$
\begin{equation}\label{labEqBohConvolution}
(\bohr{u}{v})(q):= \lim_{N \to \infty} \frac{1}{2N +1} \sum_{|l| \leq N} u(l) v(q-l).
\end{equation}
In this case, the convolution is again a complex valued  function with domain $\mathbb{Z}$. It will be useful to introduce the following  notation for the partial sums in the Bohr convolution:
\begin{equation}\label{labEqBohConvolutionpartial}
(\bohrp{u}{v})(q):=  \frac{1}{2N +1} \sum_{|l| \leq N} u(l) v(q-l).
\end{equation}
Hence,
\[
(\bohr{u}{v})(q)= \lim_{N \to \infty}(\bohrp{u}{v})(q).
\]

\section{Continuous observation of the price's path}\label{labsec:continuousobservation}
A fundamental observation in the papers  \cite{Malliavin2002} and \cite{Malliavin2009} is that the system of Fourier coefficients $\fourier[\vol]$ computed with information in the non-observable spot volatility $\vol$ decomposes into two parts, a first part  given by  $\fourierb[\vol]$, defined below,  which is obtained through Bohr's convolution of complex-valued functions constructed with `observable information' provided by  $dH$  (rigorously through pathwise It\^o integral), and a second  part that under general conditions is a ``remainder'', in that, under suitable conditions it converges to zero for $N \to \infty$. We formulate this as \thref{labth:fundamentalobservation} below due to its fundamental importance.  Before that, we introduce notation. For $q \in \mathbb{Z}$ we define the process $\Gamma(q)$ on $[-\pi,\pi]$ by 
\[
\Gamma_z(q)=\Gamma_z[dH](q):= \frac{1}{2 \pi} \int_{-\pi}^{z}  e^{-\imath q t} dH_t,\quad \Gamma_{-\pi}(q)=0, \quad z \in [-\pi,\pi].
\]

\begin{definition}
Let 
\[
\fourier[dH](q):= \Gamma_{\pi}(q).
\]
The system of coefficients $\{\fourierb_{N}[\vol]\}_{N \in \nn}$ is defined by 
\[
\fourierb_{N}[\vol](q) := 2 \pi \left\lbrace \bohrp{\fourier[dH]}{\fourier[dH]} \right\rbrace (q), \quad q \in \mathbb{Z}.
\]	
Now we define the Fourier estimator of Malliavin and Mancino under continuous observation. The Fourier estimator  of $\vol$  is defined by 
\begin{equation}\label{labEq:FejerexpressionContinuous}
\trigb_{N,M}[\vol](t):=\sum_{|l|\leq M}\left(1-\frac{|l|}{M} \right) \fourierb_{N}[\vol](l) e^{\imath l t}.
\end{equation}
\end{definition}

After preliminary notation now we have the following fundamental result.
\begin{proposition}\thlabel{labth:fundamentalobservation}
\begin{equation}\label{labeq:appbybohrconvo}
\fourier[\vol](q) = \fourierb_{N}[\vol](q)-  \resi,
\end{equation}
where
\begin{equation}
\resi[t]:=\frac{2\pi}{2N +1}\sum_{|l| \leq N} \left\lbrace \int_{-\pi}^{t}\Gamma_z(q-l) d\Gamma_z(l) + \int_{-\pi}^{t}\Gamma_z(l) d\Gamma_z(q-l)\right\rbrace.
\end{equation}
\end{proposition}
Equation \eqref{labeq:appbybohrconvo} suggests that the Fourier coefficients $\fourier[\vol]$ can be approximated by the sequence of Bohr convolutions $\{\fourierb_N[\vol]\}_{N \in \nn}$. This has  been proved by \cite{Malliavin2009} under suitable integrability conditions.  In this section we prove  that $\resi$ converges to zero a.s., generalizing this approximation to our setting. We give convergence rates  that allow us to obtain a uniform  convergence, in a precise sense, for coefficients of order $q$ in a band of the form $|q| \leq N$. From this uniform convergence we derive that the corresponding trigonometric polynomial with Ces\`aro means of coefficients  $\fourierb_N[\vol]$  converges uniformly to the corresponding polynomial constructed from the coefficients  $\fourier[\vol]$; see \thref{labthm:PathUC} below.
\subsection{Error's representation}\label{labsec:errorrepresentation}
We denote by $\tilde D$ the ``rescaled'' Dirichlet kernel $\frac{1}{2N+1}D_N$; see \eqref{labEq:RescaledDirichletKernel}.  In this section we give a convenient representation for the remainder $\resi$. 
For $t \in [-\pi, \pi]$, define
\begin{align*}
\xi_z(q,N,t)&:=\int_{-\pi}^{z} e^{-\imath q {s}}  \tilde D_N({t}-{s}) dH_{s}, \quad \xi_{-\pi}(q,N,t)=0,  \quad z \in [-\pi,t],\\
Y_t(q,N)&:= \int_{-\pi}^{t} \xi_s(q,N,s)dH_{s}, \quad Y_{-\pi}(q,N)=0,\\
Z_t(q,N)&:= \int_{-\pi}^{t} e^{-\imath q {s}} dY_{s}(0,N), \quad Z_{-\pi}(q,N)=0.
\end{align*}

The proof of the next result  is straightforward and we omit it.
\begin{lemma}\thlabel{labLemmaUpperbound}
We have that 
\begin{align*}
2 \pi \resi
&=
\int_{-\pi}^{\pi} dH_z \int_{-\pi}^{z} e^{-\imath q s} \tilde D_N(s-z) dH_s + \int_{-\pi}^{\pi} e^{-\imath q z}  dH_z \int_{-\pi}^{z} \tilde D_N(z-s) dH_s
\\
&=Y_{\pi}(q,N) + Z_{\pi}(q,N).
\end{align*}
\end{lemma}
\subsection{A key estimation: The remainder's $\mathbb{L}_p$-norm}\label{labsec:keyestimation}
Let $p > 2$  and $\alpha,\beta \in (1,\infty)$ with $\frac{1}{\alpha} +\frac{1}{\beta}=1$.  In this section we assume that $\sigma$ satisfies  \thref{lab:integrabilityforsigma} with exponent $\integrabilityratea = p(\alpha \vee \beta)$. In this case, the following constant is finite:
\begin{equation}\label{labeq:Kpab}
\left( \tilde K_{p,\alpha,\beta}\right)^p := 2^p C_p (2 \pi)^{\frac{p}{2}-1}  \sqrt[\beta]{2\pi  C_{p \beta}} \sqrt[\alpha]{E\left[\int_{-\pi}^{\pi}\sigma^{p \alpha}_s ds\right] } 
\sqrt[\beta]{ E\left[\left(\int_{-\pi}^{\pi} \sigma^{2\alpha}_s ds\right)^{\frac{p \beta}{2 \alpha}} \right]},
\end{equation}
where $C_p$ and $C_{\beta p}$ are the constants in the BDG inequality with the obvious exponents, and $B_{2\beta}$ is the constant in \thref{labLemma:DirichletkernelEstimation} below for $r=2\beta$.\\

Recall the norm $\triplenorm{\cdot}$ for a complex-valued process in \eqref{labeq:triplenorm}.
\begin{theorem}\thlabel{labthmest1unbounded}
Let $p>2$  and $\alpha,\beta \in (1,\infty)$ with $\frac{1}{\alpha} +\frac{1}{\beta}=1$. If $\sigma$ satisfies the \thref{lab:integrabilityforsigma} with exponent $\integrabilityratea = p(\alpha \vee \beta)$,  then
\begin{align}
\triplenorm{Y(q,N)}\leq
\tilde K_{p,\alpha,\beta} B_{2 \beta} N^{-\frac{1}{2 \beta}},\label{labal:upperboundY}\\
\triplenorm{Z(q,N)}\leq
\tilde K_{p,\alpha,\beta} B_{2 \beta} N^{-\frac{1}{2\beta}}.\label{labal:upperboundZ}
\end{align}
where  $B_{2\beta}$ is the constant in \thref{labLemma:DirichletkernelEstimation} for $r=2 \beta$.
Hence:
\begin{equation} \label{labal:upperboundresi}
\triplenorm{\resi[]}
\leq 
\tilde K_{p,\alpha,\beta} B_{2 \beta} N^{-\frac{1}{2 \beta}}.
\end{equation}
\end{theorem}
\begin{proof}
Note that  $2 \pi \triplenorm{\resi[]} \leq \triplenorm{Y(q,N)} + \triplenorm{Z(q,N)}$ so it suffices to prove \eqref{labal:upperboundY} for $Y(q,N)$, respectively \eqref{labal:upperboundZ} for $Z(q,N)$ in order to establish \eqref{labal:upperboundresi}. The proof for $Z(q,N)$ is similar to that of $Y(q,N)$ so we omit it. Moreover, we have $\triplenorm{Y(q,N)} \leq \triplenorm{\Re Y(q,N)} + \triplenorm{\Im Y(q,N)}$. The construction of an upper bound for $\triplenorm{\Im Y(q,N)}$ is similar to that of $\triplenorm{\Re Y(q,N)}$  so we also omit it.  Thus, we focus on the estimation of $\triplenorm{\Re Y(q,N)}$.\\	
	
For $p >2$
\begin{align*}
	E\left[\left(  \sup_{-\pi\leq s \leq \pi} |\Re Y_s(q,N)|\right)^p \right]
	& \leq 
	C_p E\left[\left\langle  \int_{-\pi}^{\cdot} \Re \xi_z(q,N,z)dH_{z}\right\rangle ^{\frac{p}{2}}_{\pi} \right]\\
	&=C_p E\left[\left( \int_{-\pi}^{\pi} \left|  \Re \xi_z(q,N,z)\right|^2 |\sigma_z|^2 dz\right) ^{\frac{p}{2}} \right],
\end{align*}
where the  inequality holds true by the BDG-inequality for a positive constant $C_p$. Moreover
\begin{align*}
	&E\left[\left( \int_{-\pi}^{\pi}|\sigma_z|^2 \left|  \Re \xi_z(q,N,z)\right|^2 dz\right)^{\frac{p}{2}} \right]\\
	& \leq 
	(2\pi)^{\frac{p}{2}-1}\sqrt[\alpha]{E\left[{\int_{-\pi}^{\pi}\left| \sigma_z\right|^{p\alpha} dz}\right]}
	\sqrt[\beta]{E\left[{\int_{-\pi}^{\pi} \left|  \Re \xi_z(q,N,z)\right|^{p \beta} dz}\right]}\\
	& = (2 \pi)^{\frac{p}{2}-1} \sqrt[\alpha]{E\left[{\int_{-\pi}^{\pi}\left| \sigma_z\right|^{p\alpha} dz}\right]} \sqrt[\beta]{\int_{-\pi}^{\pi}  E\left[ \left|  \Re \xi_z(q,N,z)\right|^{p{\beta}}\right] dz},
\end{align*}
where the  inequality is obtained from H\"older and Jensen inequalities, and the last equality holds true by Tonelli-Fubini's Theorem.\\
	
Note that for $t \in [-\pi,z]$
\begin{align*}
\Re \xi_{t}(q,N,z) 
&= 
\int_{-\pi}^{t} \left( \Re e^{-\imath q {s}} \right)  \tilde D_N(z-s) dH_{s}
= 
\int_{-\pi}^{t} \cos(qs) \tilde D_N(z-s) dH_{s}.
\end{align*}
Hence,
\begin{align*}
E\left[ \sup_{-\pi \leq t \leq z}\left|  \Re \xi_t(q,N,z)\right|^{p\beta}\right]
&\leq C_{p\beta} B_{2\beta}^{p\beta} E\left[\left( \int_{-\pi}^{\pi} |\sigma_s|^{2 \alpha} ds \right)^{\frac{p\beta}{2 \alpha}} \right] \frac{1}{N^{\frac{p}{2}}},
\end{align*}
due to \thref{lablem:auxestiH}  with $\kappa=p \beta$.\\

Combining these estimates:
\begin{align*}
E \left[  \left( \sup_{-\pi \leq s \leq \pi} |\Re Y_s(q,N)|\right)^p\right]
&\leq  C_p (2 \pi)^{\frac{p}{2}-1} \sqrt[\alpha]{E\left[\int_{-\pi}^{\pi}|\sigma_z|^{p \alpha} dz\right]} 
\sqrt[\beta]{2\pi  C_{p \beta}  B_{2 \beta}^{p \beta}   E\left[\left(\int_{-\pi}^{\pi} |\sigma_s|^{2\alpha} ds\right)^{\frac{p \beta}{2 \alpha}} \right] \frac{1}{N^{\frac{p}{2}}}}\\
&=C_p (2 \pi)^{\frac{p}{2}-1}B_{2 \beta}^{p}  \sqrt[\beta]{2\pi  C_{p \beta}} \sqrt[\alpha]{E\left[\int_{-\pi}^{\pi} |\sigma_z|^{p \alpha} dz\right]} 
\sqrt[\beta]{ E\left[\left(\int_{-\pi}^{\pi} |\sigma_s|^{2\alpha} ds\right)^{\frac{p \beta}{2 \alpha}} \right]}
\frac{1}{N^{\frac{p}{2 \beta}}}.
\end{align*}
Hence
\begin{align*}
\sqrt[p]{E \left[  \left( \sup_{-\pi \leq s \leq \pi} |\Re Y_s(q,N)|\right)^p\right]}
& \leq \frac{1}{2} B_{2 \beta} \tilde K_{p,\alpha,\beta}\frac{1}{N^{\frac{1}{2 \beta}}}.
\end{align*}
Adding up the imaginary part we obtain  \eqref{labal:upperboundY} for $Y$. The proof of \eqref{labal:upperboundZ} is similar. The inequality \eqref{labal:upperboundresi} is a consequence of \eqref{labal:upperboundY} and \eqref{labal:upperboundZ}, due to \thref{labLemmaUpperbound}, since $\triplenorm{\cdot}$ satisfies the triangle inequality.
\end{proof}
\begin{remark}
A simpler formulation of \thref{labthmest1unbounded} is that, for any exponent $r\in (0,\frac{1}{2})$, given  sufficient integrability of $\sigma$ depending on $r$, 
\begin{equation*}
\triplenorm{\resi[]}=O\left( N^{-r}\right).
\end{equation*}
Keeping track of the constant $\tilde K_{p,\alpha,\beta}$ exhibits explicit dependence on the exponent $p$ in the constant appearing in the big O notation $O(\cdot)$.
\end{remark}
\subsection{Coefficients' convergence} \label{labsec:coefficientconvergence}
The notation for the running supremum of a complex valued process was introduced in equation \eqref{labeq:runningsupremumcomplexc}. For the process $\resi[]$:
\begin{align*}
\resi^*&=\sup_{-\pi \leq s \leq \pi} |\Re \resi[s]| + \sup_{-\pi \leq s \leq \pi} |\Im \resi[s]|.
\end{align*}

We adopt the notation and conditions of \thref{labthmest1unbounded}. Fix $\pow \in (0,\frac{1}{2 \beta})$. Given $M \in \nn$, the  `good event' of a Small Error is defined by:
\begin{align*}
SE(N,M)
&:=\bigcap_{|q| \leq M} \left\lbrace \resi^*  < 6 B_{2 \beta} {\tilde K_{p,\alpha,\beta}}N^{\pow -\frac{1}{2\beta}}\right\rbrace\\ &=\left\lbrace \sup_{|q| \leq M} \resi^*   < 6 B_{2 \beta} {\tilde K_{p,\alpha,\beta}}N^{\pow -\frac{1}{2\beta}}\right\rbrace.
\end{align*}
The complement of $SE(M,N)$ in which a Large Error is possible is given by
\[
LE(N,M):= \Omega \setminus SE(N,M)=\bigcup_{|q| \leq M} \left\lbrace \resi^* \geq 6 B_{2 \beta} {\tilde K_{p,\alpha,\beta}}N^{\pow -\frac{1}{2\beta}}\right\rbrace.
\]
 
\begin{lemma}\thlabel{lablem:probGoodEvent}
Take the notation and conditions of \thref{labthmest1unbounded}. For  $\pow \in (0,\frac{1}{2 \beta})$ we have
\begin{align*}
P\left(SE(N,M) \right) 
& \geq 1 - (2M+1)\frac{1}{N^{\pow p}}.
\end{align*}
\end{lemma}
\begin{proof}
Let $\mu:=E\left[ \resi^*\right]$. For $p$ with $p \geq 2$, we have 
\begin{align*}
	\sqrt[p]{E\left[\left| \resi^* - \mu \right|^p \right] } 
	&\leq  2 \triplenorm{\resi[]}\\
	& \leq 
	2 B_{2 \beta} {\tilde K_{p,\alpha,\beta}} N^{-\frac{1}{2 \beta}},
\end{align*}
where the first inequality follows from the triangle inequality, and the second from the estimation in \thref{labthmest1unbounded}.

For $c>1$, the set inclusions
\begin{align*}
\left\lbrace \resi^* \geq c 6 B_{2 \beta} {\tilde K_{p,\alpha,\beta}} N^{-\frac{1}{2\beta}}\right\rbrace  
&\subseteq 
\left\lbrace \resi^* \geq \mu + c \sqrt[p]{E\left[\left|  {\resi^*} - \mu \right|^p \right] } \right\rbrace\\ 
&\subseteq \left\lbrace |\resi^*-\mu| \geq c \sqrt[p]{E\left[\left|\resi^* - \mu\right|^p \right] }\right\rbrace 
\end{align*}
yield
\[
P\left(\left\lbrace {\resi^*} \geq c 6  B_{2 \beta} {\tilde K_{p,\alpha,\beta}} N^{-\frac{1}{2\beta}}\right\rbrace   \right) \leq  \frac{1}{c^p},
\]
due to Chebyshev's inequality for higher moments; see \thref{labpro:Chebyshev} below. In particular, for $c=N^{\pow }$ with $\pow \in (0,\frac{1}{2\beta})$
\[
P\left(\left\lbrace \resi^* \geq 6 B_{2 \beta} {\tilde K_{p,\alpha,\beta}} N^{\pow -\frac{1}{2\beta}}\right\rbrace  \right) \leq  \frac{1}{N^{\pow p}}.
\]
As a consequence, the  event $LE(N,M)$ has probability less than $(2M+1)N^{-\pow p}$ and
by taking set complement the result follows.
\end{proof}

\begin{lemma}\thlabel{lablem:ucerrorhighprob}
Take $\pow \in (0,\frac{1}{2\beta})$,  $p > 2$ and $r >0$. If $\pow p-r>1$ and the conditions of \thref{labthmest1unbounded} hold true, then, for $M=O(N^r)$
\[
P(LE(N,M))=O(N^{r-\pow p}).
\]
Moreover, in the complement of a null event
\begin{align}\label{labal:eventualshrinkrate}
\sup_{|q| \leq M}  \resi^* < 6 B_{2 \beta} {\tilde K_{p,\alpha,\beta}} N^{\pow -\frac{1}{2\beta}}, \text{ eventually}.
\end{align}
\end{lemma}
\begin{proof}
The event $LE(N,M)$ has probability less than $(2M+1)N^{-\pow p}$ due to \thref{lablem:probGoodEvent}. For $M=O(N^r)$ we have $P(LE(N,M)) =O(N^{r-\pow p})$. Indeed, for  $M \leq K N^r$
\begin{align*}
P(LE(N,M)) \leq  (2KN^r+1)\frac{1}{N^{\pow p}} \leq 2K \frac{1}{N^{\pow p-r}} +  \frac{1}{N^{\pow p}}\leq \frac{2K+1}{N^{\pow p -r}}.
\end{align*}
As a consequence
\[
\limsup_{M=O(N^r), N \to \infty} LE(N,M)
\]
is a null event, due to the  Borel-Cantelli lemma.  
\end{proof}
\begin{remark}
In the second part of \thref{lablem:ucerrorhighprob} the qualifier ``eventually'' means that for $\omega \in \tilde \Omega$, where $\tilde \Omega$ is an event with full measure, there exists $I(\omega) \in \nn$ such that for any $N \geq I(\omega)$ the estimation \eqref{labal:eventualshrinkrate} holds true.  How large must $I(\omega)$ be?. Selecting a measurable index, we can estimate its tail behavior as follows.   Let $K>0$ be such that $M(N) \leq KN^r$. The event $\tilde \Omega$ can be taken as 
\begin{align*}
\tilde \Omega
&= \bigcup_{j=1}^{\infty} \bigcap_{N=j}^{\infty}SE(N,M(N)).
\end{align*}
For a concrete index define $I(\omega)=\inf\{j \in \nn \mid \omega \in SE(n, M(n)) \text{ for all } n \geq j\}$.  Note that $\bigcap_{N=j}^{\infty}SE(N,M(N)) \subset \{I \leq j\}$ or equivalently $\{I>j\} \subset \bigcup_{N=j}^{\infty}LE(N,M(N))$. Hence
\begin{align*}
P\left(\{I >j\} \right) &  \leq \sum_{N=j}^{\infty} P\left(LE(N,M(N)) \right)\\
&\leq \sum_{N=j}^{\infty} \frac{2K+1}{N^{\pow p-r}}\\
&\leq  \frac{2K+1}{\pow p -r -1}(j-1)^{1- \pow p + r}.
\end{align*}
As a consequence,  $P\left(\{I >j\} \right) =O((j-1)^{1- \pow p + r})$.
\end{remark}

\begin{theorem}\thlabel{labthm:coeffConvergence}
Take $\pow \in (0,\frac{1}{2\beta})$,  $p \geq 2$ and $r >0$. If $\pow p-r>1$  and the conditions of \thref{labthmest1unbounded} hold true, then, for $M=O(N^r)$
\[
P\left(\left\lbrace \sup_{|q| \leq M}  \left| \fourier[\vol] (q)-\fourierb_{N}[\vol](q)\right| 
\geq
6 B_{2 \beta} {\tilde K_{p,\alpha,\beta}} N^{\pow -\frac{1}{2\beta}} \right\rbrace \right) = O\left(\frac{1}{N^{\pow p -r}}\right).
\]
Hence, on the complement of a null event 
\begin{align*}
\sup_{|q| \leq M} \left\lbrace \left| \fourier [\vol](q)-\fourierb_{N}[\vol](q)\right| \right\rbrace   < 6 B_{2 \beta} {\tilde K_{p,\alpha,\beta}} N^{\pow -\frac{1}{2\beta}}, \quad \text{eventually}.
\end{align*}
In particular
\[
\lim_{N\to \infty} \fourierb_{N}[\vol](q)= \fourier[\vol](q), \quad a.s.
\]
\end{theorem}
\begin{proof}
This is a direct consequence of \thref{lablem:ucerrorhighprob} since
\[
\left| \fourier[\vol] (q)-\fourierb_{N}[\vol] (q)\right|\leq \resi^*.
\]
\end{proof}

\subsection{Uniform convergence of trigonometric polynomials}\label{labsec:polynomialconvergence}
Let $\trig_M[\vol]$ be the trigonometric polynomial of $\vol$ determined by the system of `exact coefficients' $\fourier[\vol]$, that is
\begin{equation}\label{labeq:trigvolexaccoeff}
\trig_{M}[\vol](t):=\sum_{|l|\leq M}\left(1-\frac{|l|}{M} \right) \fourier[\vol](l) e^{\imath l t}.
\end{equation}
The next theorem establishes that the trigonometric polynomial $\trigb_{N,M}[\vol]$ defined in \eqref{labEq:FejerexpressionContinuous}  with coefficients $\fourierb_N[\vol]$  
constructed with the Bohr convolution is an accurate approximation comparable with the trigonometric approximation \eqref{labeq:trigvolexaccoeff} determined by  the `exact coefficients' $\fourier[\vol]$ under an appropriate growth regime for $M$, the order of the trigonometric polynomials.
\begin{theorem}\thlabel{labthm:PathUC}
Take $\pow \in (0,\frac{1}{2\beta})$,  $p >2$ and $r >0$. If $r<(\frac{1}{2\beta} - \pow)\wedge (\pow p-1)$ and the conditions of \thref{labthmest1unbounded} hold true, then, for $M=O(N^r)$
\begin{equation}\label{labeq:uniformconvergenceofapproxPoly}
\lim_{\substack{N,M\to \infty \\ M=O(N^r)}}\sup_{t \in [-\pi,\pi]}	\left|\trig_{M}[\vol](t) - \trigb_{N,M}[\vol](t)\right| = 0, a.s.
\end{equation}
\end{theorem}
\begin{proof}
Recall the definition of $\trigb_{N,M}[\vol]$ in \eqref{labEq:FejerexpressionContinuous}. Assume that $M \leq K N^r$. For $t \in [-\pi,\pi]$ 
\begin{align*}
\left|\trig_{M}[\vol](t) - \trigb_{N,M}[\vol](t)\right|
&=
\left|\sum_{|l|\leq M}\left(1-\frac{|l|}{M} \right) \left\lbrace  \fourier[\vol](l) - \fourierb_N[\vol](l) \right\rbrace  e^{\imath l t}\right|\\
& \leq
M \sup_{|l| \leq M}   \left|  \fourier[\vol](l) - \fourierb_N[\vol](l) \right|\\ 
& \leq M 6 B_{2 \beta} {\tilde K_{p,\alpha,\beta}} N^{\pow -\frac{1}{2\beta}}  \textit{ eventually}\\
& \leq K 6 B_{2 \beta}  {\tilde K_{p,\alpha,\beta}} N^{r+\pow -\frac{1}{2\beta}},
\end{align*}
where the second inequality holds true eventually in the complement of a null event due to \thref{labthm:coeffConvergence}, since $\pow p-r>1$.
Thus, \eqref{labeq:uniformconvergenceofapproxPoly} holds true. 
\end{proof}

\begin{remark}
Flexibility in the  exponents. In \thref{labthm:PathUC} the exponent $r$ is strictly less than $\frac{1}{2}$. There is a trade-off for this exponent. Selecting $r$ close to  $\frac{1}{2}$ allows a better growth rate for $M$, the number of harmonics in the trigonometric approximation. However, a larger $r$ worsens the rate at which the trigonometric polynomial $\trigb_{N,M}[\vol]$  constructed with coefficients estimated through Bohr convolution approximates  the trigonometric polynomial $\trig_{M}[\vol]$ constructed with `exact coefficients'. Hence, an optimal choice lies in some intermediate value.

Taking $r$ to its upper bound requires: $\beta \searrow 1$, $\pow \searrow 0$ and  $p \nearrow \infty$. An exponent $\beta$ closer to its lower bound requires an exponent $\alpha$ arbitrarily large. Thus,  demanding more integrability on the paths of $\vol$. As it was mentioned in \thref{labrem:easydiffusions}, diffusions with Lipschitz coefficients and quadratic growth satisfy this requirement; see \thref{labtheasydiffusionscorollaryintegrability}.
Thus, in specific models where integrability of $\sigma$ with a high exponent is available, more flexibility is gained for a better choice of the exponents. 
\end{remark}

\begin{remark}\thlabel{labrem:moduluscontinuity}
An important property of the Fourier estimator of volatility is its capability of approximating the whole path $\{\vol_{t}(\omega)\}_{-\pi \leq t \leq \pi}$ in a given scenario $\omega$. In this regard, 
\thref{labthm:PathUC} shows  that the trigonometric polynomial $\trigb_{N,M}[\vol]$ will approximate the volatility $\vol$ with similar precision as the trigonometric polynomial $\trig_{M}[\vol]$. This in turn will be the case in our present setting of a volatility with  continuous paths. The approximation is uniform  in compact intervals contained in  $(-\pi,\pi)$. A quantitative formulation can be obtained from the modulus of continuity of $\vol$; see e.g., \cite[Corollary 1.6.5 p.82]{Butzer1971}. 
A general class of diffusions provides concrete instances of this uniform approximation; see \thref{labtheasydiffusionscorollarymodulusofcontinuity}.
\end{remark}

\section{Discrete observation}\label{labsec:discreteobservation}
Take a family of partitions $\{\nu^n\}_{n \in \nn}$ of the interval $[-\pi,\pi]$ with 
\[
\nu^n=\{t_0^n=-\pi \leq t_1^n \leq \ldots \leq t_{m_n}^n=\pi \}.
\]
Let $\rho(n):=\sup_{i=0,\ldots,m_n-1}\{|t^n_{i+1}-t^n_{i}|\}$ be the norm of the partition $\nu_n$ and assume that $\rho(n)\to 0$.\\ 

In this section we develop the analogous results to those in Section \ref{labsec:continuousobservation} under  a discrete observation of the price process along the system of partitions $\{\nu^n\}_{n \in \nn}$. The steps and proofs to follow are  similar and therefore, we only give the statements of results and the parts of the proofs that need to be modified.\\

For the partition $\nu^n$ we define a function that selects for a given time $t \in [-\pi,\pi]$ the closest element from the left of the partition:  
\begin{align}\label{labeleq:projectionTOpartition}
\cut{t}&:=\sup\{z \in \nu^n \mid z \leq t\}.
\end{align}
For $q \in \mathbb{Z}$ we define the process $\Gamma(q,n)$ by 
\[
\Gamma_{z}(q,n)=\Gamma_{z}[dH](q,n):= \frac{1}{2 \pi} \int_{-\pi}^{z}  e^{-\imath q \cut{t}} dH_t, \quad \Gamma_{-\pi}(q,n)=0,  \quad z \in [-\pi,\pi].
\]
\begin{definition}
Let 
\begin{align}\label{labEq:discretizedDirichletKernel}
\fourier_n[dH](q)&:=\Gamma_{\pi}(q,n), \quad q \in \mathbb{Z}.
\end{align}
The convolution system of Fourier coefficients $\{\fourierb_{n,N}[\vol](q)\}_{q \in \mathbb{Z}}$ of the volatility $\vol$ under discrete observation  is defined by:
\begin{equation}
\fourierb_{n,N}[\vol](q):= 2\pi \left\lbrace  \bohrp{\fourier_n[dH]}{\fourier_n[dH]}\right\rbrace(q) .  	
\end{equation}
The Fourier estimator  of the spot volatility under discrete observation is 
\begin{equation}\label{labEq:FejerexpressionDiscretized}
\trigb_{n,N,M}[\vol](t):=\sum_{|l|\leq M}\left(1-\frac{|l|}{M} \right) \fourierb_{n,N}[\vol](l) e^{\imath l t}.
\end{equation}
\end{definition}
It is clear that the convolution system of Fourier coefficients is an  ``approximate system'' and it is known from the papers \cite{Malliavin2002} and \cite{Malliavin2009} that the error in the approximation is influenced by two procedures: truncating a series and replacing an integral by a sum. First, it is an approximation from the fact that coefficients are constructed from a truncated series in the Bohr convolution, and second,  discrete observation of $H$ yields approximate estimation of integrals.\\

The estimator $\trigb_{n,N,M}[\vol]$ is the discretized version of \eqref{labEq:FejerexpressionContinuous}. The next result is the discretized version of \thref{labth:fundamentalobservation}.  We omit the proof.

\begin{proposition}\thlabel{labth:fundamentalobservationdiscretized}
We have
\begin{equation}\label{labeq:appbybohrconvodiscretized}
\fourier_{n}[\vol](q) = \fourierb_{n,N}[\vol](q)- \residis,
\end{equation}
where
\begin{equation}
\residis:=\frac{2\pi}{2N +1}\sum_{|l| \leq N} \left\lbrace \int_{-\pi}^{\pi}\Gamma_t(q-l,n) d\Gamma_t(l,n) + \int_{-\pi}^{\pi}\Gamma_t(l,n) d\Gamma_t(q-l,n)\right\rbrace.
\end{equation}
\end{proposition}

\subsection{Error representation} \label{labsec:errorrepresentationdiscrete}
For $t \in [-\pi ,  \pi]$, define
\begin{align*}
\xi_z(q,n,N,t)&:=\int_{-\pi}^{z} e^{-\imath q \cut{s}}  \tilde D_N(\cut{t}-\cut{s}) dH_{s}, \quad \xi_{-\pi}(q,n,N,t)=0,\quad z \in [-\pi,t],\\
Y_t(q,n,N)&:= \int_{-\pi}^{t} \xi_s(q,n,N,s)dH_{s}, \quad Y_{-\pi}(q,N)=0,\\
Z_t(q,n,N)&:= \int_{-\pi}^{t} e^{-\imath q \cut{s}} dY_{s}(0,n,N), \quad Z_{-\pi}(q,n,N)=0.
\end{align*}
The next result is the discretized version of  \thref{labLemmaUpperbound} and we omit the proof. 
\begin{lemma}\thlabel{labLemmaUpperbounddiscretized}
We have  
\begin{align*}
2 \pi \residis   &= Y_{\pi}(q,n,N) + Z_{\pi}(q,n,N).
\end{align*}
\end{lemma}
\subsection{A key estimation: The  $\mathbb{L}_p$-norm of the remainder}\label{labsec:keyestimationdiscretized}
Take $p> 2$  and $\alpha,\beta \in (1,\infty)$ with $\frac{1}{\alpha} +\frac{1}{\beta}=1$.  Let $\sigma$ satisfy  \thref{lab:integrabilityforsigma} with exponent $\integrabilityratea=p(\alpha \vee \beta )$.  Recall the definition of the constant $\tilde K_{p,\alpha,\beta}$ in \eqref{labeq:Kpab}. It is given by 
\begin{equation*}
	\left( \tilde K_{p,\alpha,\beta}\right)^p := 2^p C_p (2 \pi)^{\frac{p}{2}-1}  \sqrt[\beta]{2\pi  C_{p \beta}} \sqrt[\alpha]{E\left[\int_{-\pi}^{\pi}\sigma^{p \alpha}_s ds\right] } 
	\sqrt[\beta]{ E\left[\left(\int_{-\pi}^{\pi} \sigma^{2\alpha}_s ds\right)^{\frac{p \beta}{2 \alpha}} \right]}.
\end{equation*}
We also define for $r>1$ the constant
\begin{equation}\label{labeq:Aconstant}
\dddot A_r:= 5 + \frac{2\pi^r}{r-1}.
\end{equation}

The next result is the discretized version of \thref{labthmest1unbounded}.
\begin{theorem}\thlabel{labthmest1unboundeddiscretized}
Take $p > 2$ and $\alpha,\beta \in (1,\infty)$ with $\frac{1}{\alpha} +\frac{1}{\beta}=1$. If $\sigma$ satisfies the \thref{lab:integrabilityforsigma} with exponent $\integrabilityratea=p(\alpha \vee \beta)$, then
\begin{align}
\triplenorm{Y(q,n,N)}
\leq
\tilde{K}_{p,\alpha,\beta} \left( 5 \rho(n) + \dddot{A}_{2\beta} (2N+1)^{-1} \right)^{\frac{1}{2 \beta }},\label{labal:upperboundYdiscretized}\\
\triplenorm{Z(q,n,N)}
\leq 
\tilde{K}_{p,\alpha,\beta} \left( 5 \rho(n) + \dddot{A}_{2\beta} (2N+1)^{-1} \right)^{\frac{1}{2 \beta }}.\label{labal:upperboundZdiscretized}
\end{align}
Hence:
\begin{equation} \label{labal:upperboundresidiscretized}
\triplenorm{\residis[]}\leq  \tilde{K}_{p,\alpha,\beta} \left( 5 \rho(n) + \dddot{A}_{2\beta} (2N+1)^{-1} \right)^{\frac{1}{2 \beta }}.
\end{equation}

\end{theorem}
\begin{proof}
As in the proof of \thref{labthmest1unbounded}, in order to show \eqref{labal:upperboundYdiscretized}, we only present how to obtain an upper bound for  $\left\| \sup_{-\pi \leq s \leq \pi} |\Re Y_s(q,n,N)| \right\|_{\mathbb{L}_p}$. Moreover, since the proof requires similar arguments we only sketch the main steps.\\

For $p > 2$ and $\alpha,\beta \in (1,\infty)$ with $\frac{1}{\alpha} +\frac{1}{\beta}=1$, in a similar way as in \thref{labthmest1unbounded} we get
\begin{align*}
&E\left[\left(  \sup_{-\pi \leq s \leq \pi} |\Re Y_s(q,n,N)|\right) ^p \right]
\\
\leq  &C_p (2 \pi)^{\frac{p}{2}-1} \sqrt[\alpha]{E\left[{\int_{-\pi}^{\pi}\left| \sigma_z\right|^{p\alpha} dz}\right]} \sqrt[\beta]{\int_{-\pi}^{\pi}  E\left[ \left|  \Re \xi_z(q,n,N,z)\right|^{p{\beta}}\right] dz}.
\end{align*}
	
For $z \in [-\pi, \pi]$, the expected value $E\left[\sup_{-\pi \leq t \leq z}\left|  \Re \xi_t(q,n,N,z)\right|^{p \beta}\right]$, hence
$E\left[ \left|  \Re \xi_z(q,n,N,z)\right|^{p \beta}\right]$, can be estimated from above as follows:
\begin{align*}
&E\left[\sup_{-\pi \leq t \leq z}\left|  \Re \xi_t(q,n,N,z)\right|^{p \beta}\right]\\
& \leq  C_{p \beta} E\left[\left( \int_{-\pi}^{\pi} |\sigma_s|^{2 \alpha} ds \right)^{\frac{p \beta}{2 \alpha}} \right] \left(\int_{-\pi}^{z}|\tilde D(\cut{z} - \cut{s})|^{2 \beta} ds\right)^{\frac{p \beta}{2 \beta}} \\
& \leq  C_{p \beta} E\left[\left( \int_{-\pi}^{\pi} |\sigma_s|^{2 \alpha} ds \right)^{\frac{p \beta}{2 \alpha}} \right] \left(5 \rho(n) + \dddot{A}_{2\beta} (2N+1)^{-1}\right)^{\frac{p \beta}{2 \beta}}, 
\end{align*}
where the first inequality follows from BDG and H\"older's inequalities, and the fact that $\tilde D$ is deterministic. The second inequality follows from   \thref{labLemma:DirichletkernelEstimationdiscretized}. Combining these estimates, we obtain:
\begin{align*}
&E\left[\sup_{-\pi  \leq s \leq \pi}\left|  \Re Y_s(q,n,N)\right|^{p}\right]\\ 
\leq  
&C_p (2 \pi)^{\frac{p}{2}-1} \sqrt[\alpha]{E\left[\int_{-\pi}^{\pi}|\sigma_z|^{p \alpha} dz\right]} 
\sqrt[\beta]{2\pi  C_{p \beta}  E\left[\left(\int_{-\pi}^{ \pi} |\sigma_s|^{2\alpha} ds\right)^{\frac{p \beta}{2 \alpha}} \right] \left( 5 \rho(n) + \dddot{A}_{2\beta} (2N+1)^{-1} \right)^{\frac{p}{2}} }.
\end{align*}
Hence
\begin{align*}
\sqrt[p]{E\left[\sup_{-\pi \leq s \leq  \pi}\left|  \Re Y_s(q,n,N)\right|^{p}\right]} 
& \leq \frac{1}{2}\tilde{K}_{p,\alpha,\beta} \left( 5 \rho(n) + \dddot{A}_{2\beta} (2N+1)^{-1} \right)^{\frac{1}{2 \beta }}.
\end{align*}
Adding up the imaginary part we obtain  \eqref{labal:upperboundYdiscretized} for $Y(q,n,N)$. The proof of \eqref{labal:upperboundZdiscretized} is similar. The inequality \eqref{labal:upperboundresidiscretized} is a consequence of \eqref{labal:upperboundYdiscretized} and \eqref{labal:upperboundZdiscretized}, due to \thref{labLemmaUpperbounddiscretized}.
\end{proof}

\subsection{Coefficients' convergence} \label{labsec:coefficientconvergencediscretized}
We adopt the notation and conditions of \thref{labthmest1unboundeddiscretized}. Take  $\pow \in (0,\frac{1}{2 \beta})$. We define the  `good event' of a small error:
\begin{align*}
SE(n,N,M)
&:=\bigcap_{|q| \leq M} \left\lbrace \residis^* < 6 \tilde{K}_{p,\alpha,\beta} \left( 5 \rho(n) + \dddot{A}_{2\beta} (2N+1)^{-1} \right)^{\frac{1}{2\beta}-\pow }\right\rbrace\\ 
&=\left\lbrace \sup_{|q| \leq M} \residis^* < 6 \tilde{K}_{p,\alpha,\beta} \left( 5 \rho(n) + \dddot{A}_{2\beta} (2N+1)^{-1} \right)^{\frac{1}{2\beta}-\pow}\right\rbrace.
\end{align*}
The complement of $SE(n,M,N)$ is
\[
LE(n,N,M):= \Omega \setminus SE(n,N,M)=\bigcup_{|q| \leq M} \left\lbrace \residis^* \geq 6 \tilde{K}_{p,\alpha,\beta} \left( 5 \rho(n) + \dddot{A}_{2\beta} (2N+1)^{-1} \right)^{\frac{1}{2\beta}-\pow }\right\rbrace.
\]

The next result is the discretized version of \thref{lablem:probGoodEvent}.
\begin{lemma}\thlabel{lablem:probGoodEventdiscretized}
Take the notation and conditions of \thref{labthmest1unboundeddiscretized}. For  $\pow \in (0,\frac{1}{2 \beta})$ we have
\begin{align*}
P\left(SE(n,N,M) \right)  & \geq 1- (2M+1)\left( 5 \rho(n) + \dddot{A}_{2\beta} (2N+1)^{-1} \right)^{\pow p}.
\end{align*}
\end{lemma}
\begin{proof}
Take $p$ with $p > 2$. Denote $E[\residis^*]$ with $\mu$.  We have 
\begin{align*}
\sqrt[p]{E\left[\left| \residis^* - \mu \right|^p \right] } 
&\leq
2 \triplenorm{\residis} & \leq 
2 \tilde{K}_{p,\alpha,\beta} \left( 5 \rho(n) + \dddot{A}_{2\beta} (2N+1)^{-1} \right)^{\frac{1}{2 \beta }},
\end{align*}
due to the triangle inequality, and  \thref{labthmest1unboundeddiscretized}.

For $c>1$ 
\[
P\left(\left\lbrace {\residis} \geq c6  \tilde{K}_{p,\alpha,\beta} \left( 5 \rho(n) + \dddot{A}_{2\beta} (2N+1)^{-1} \right)^{\frac{1}{2 \beta }} \right\rbrace   \right) \leq  \frac{1}{c^p},
\]
due to Chebyshev's inequality for higher moments; see \thref{labpro:Chebyshev} below. In particular, for $c=\left( 5 \rho(n) + \dddot{A}_{2\beta} (2N+1)^{-1} \right)^{-\pow}$ with $\pow \in (0,\frac{1}{2\beta})$
\[
P\left(\left\lbrace \residis^* \geq  6  \tilde{K}_{p,\alpha,\beta} \left( 5 \rho(n) + \dddot{A}_{2\beta} (2N+1)^{-1} \right)^{\frac{1}{2\beta}-\pow} \right\rbrace \right) 
\leq  \left( 5 \rho(n) + \dddot{A}_{2\beta} (2N+1)^{-1} \right)^{\pow p}.
\]
As a consequence, the  event $LE(n,N,M)$ has probability less than $(2M+1)\left( 5 \rho(n) + \dddot{A}_{2\beta} (2N+1)^{-1} \right)^{\pow p}$ and by taking the set complement the result follows.
\end{proof}

The next result is the discretized version of \thref{lablem:ucerrorhighprob}. The proof is similar and we omit it.
\begin{lemma}\thlabel{lablem:ucerrorhighprobdiscretized}
Take $\pow \in (0,\frac{1}{2\beta})$,  $p > 2$ and $r >0$ with $\pow p -r>1$. If the conditions of \thref{labthmest1unboundeddiscretized} hold true, then, for $M=O\left( \left(5 \rho(n) + \dddot{A}_{2\beta} (2N+1)^{-1} \right)^{-r}\right)$
\[
P(LE(n,N,M))=O\left( \left(5 \rho(n) + \dddot{A}_{2\beta} (2N+1)^{-1} \right)^{\pow p-r} \right).
\]
If along a sequence $N=N_n$
\begin{equation}\label{labeq:finiteseries}
\sum  \left(5 \rho(n) + \dddot{A}_{2\beta} (2N+1)^{-1} \right)^{\pow p-r} < \infty,
\end{equation}
then on the complement of a null event
\begin{align}\label{labal:eventualshrinkratediscretized}
\sup_{|q| \leq M} \residis^*  < 6 \tilde{K}_{p,\alpha,\beta} \left( 5 \rho(n) + \dddot{A}_{2\beta} (2N+1)^{-1} \right)^{\frac{1}{2\beta}-\pow}, \quad \text{eventually}.
\end{align}
\end{lemma}
The next result is the discretized version of \thref{labthm:PathUC}. The proof is similar and we omit it.
\begin{theorem}\thlabel{labthm:PathUCdiscretized}
Take $\pow \in (0,\frac{1}{2\beta})$,  $p > 2$ and $r >0$. If $r<(\frac{1}{2\beta}-\pow)\wedge (\pow p-1)$ and the conditions of \thref{lablem:ucerrorhighprobdiscretized} hold true, then, for $M=O\left( \left( 5 \rho(n) + \dddot{A}_{2\beta} (2N+1)^{-1} \right)^{-r} \right)$
\begin{equation}\label{labeq:uniformconvergenceofapproxPolydiscretized}
\lim_{\substack{n,N\to \infty \\ M=O\left(\left( 5 \rho(n) + \dddot{A}_{2\beta} (2N+1)^{-1} \right)^{-r} \right)}}\sup_{t \in [-\pi,\pi]}	\left|\trig_{M}[\vol](t) - \trigb_{n,N,M}[\vol](t)\right| = 0, a.s.
\end{equation}
\end{theorem}

Under the assumption of the growth regime $\rho(n) N \to 0$, the condition $\pow p -r>1$ is sufficient for the general condition \eqref{labeq:finiteseries} in \thref{lablem:ucerrorhighprobdiscretized}. Hence,     \thref{labthm:PathUCdiscretized} has the following corollary.
\begin{corollary}
Take $\pow \in (0,\frac{1}{2\beta})$,  $p > 2$ and $r >0$. Assume $r<(\frac{1}{2\beta}-\pow)\wedge (\pow p-1)$ and  $\rho(n) N \to 0$. Then for $M=O\left( N^{r} \right)$
\begin{equation}\label{labeq:uniformconvergenceofapproxPolydiscretizedcor}
\lim_{\substack{N\to \infty \\ M=O\left(N^{r} \right)}}\sup_{t \in [-\pi,\pi]}	\left|\trig_{M}[\vol](t) - \trigb_{n,N,M}[\vol](t)\right| = 0, a.s.
\end{equation}
\end{corollary}
\section{Continuous observation of a c\`adl\`ag path} \label{labsec:continuousobservationJumps}
Recall that $H$  satisfies the SDE \eqref{labEqdyn}:
\[
dH_t = \sigma_t dW_t, \text{ with } H_{-\pi}=x.
\]
In this section, instead of $H$ we consider a process $\price$ that also includes jumps:
\begin{equation}\label{labEqdynJumps}
d \price_t =dH_t + dJ_t, \text{ with } \price_{-\pi}=x.
\end{equation}
where $J$ satisfies the conditions of the next assumption with  exponents specified below. 
\begin{assumption}\thlabel{labass:localpJumpsummability}
The stochastic process $J$ is a purely discontinuous local martingale and for $\integrabilityratejumpsa>1$
\begin{equation}\label{labasseq3:localpJumpsummability}
E \left[ \left(\sum_{z \in [-\pi, \pi]}  \Delta J^2_z \right)^{\integrabilityratejumpsa}  \right] <\infty.
\end{equation}

For $t \in [-\pi,\pi]$ and $\delta\in (0,\pi)$ let $M_t(\delta)$ be defined by
\begin{equation}\label{labasseq:localjumps}
M_t(\delta):= \int_{(t-\delta,t+\delta)\cap [-\pi, \pi]\setminus\{t\}}d[J]_z  = \sum_{\substack{z \in [-\pi, \pi]\setminus\{t\}\\ 0 < |t-z| < \delta}}\Delta J_z^2.
\end{equation}
There exists an exponent $\integrabilityratejumps>0$,  independent of $t$,  such that 
\begin{equation}\label{labasseq:localpJumpsummability}
\left\|M_t(\delta)\right\|_{L_{\integrabilityratejumpsa}}=
O\left( \delta^{\integrabilityratejumps}\right).
\end{equation}

Let $\{\tau_n\}_{n \in \nn}$ be a sequence of stopping times taking values in $[-\pi,\pi]\cup\{\infty\}$ that exhaust the jump times of $J$. We have
\begin{equation}\label{labasseq2:localpJumpsummabilitylefttaun}
\left\| \int_{(\tau_n -\delta,\tau_n)\cap [-\pi, \pi]} d[J]_z \right\|_{L_{\integrabilityratejumpsa}} =O(\delta^{\integrabilityratejumps}).
\end{equation}
\end{assumption}
\begin{remark}
Processes satisfying \thref{labass:localpJumpsummability} include processes with jumps of finite activity. In Section \ref{labsubs:cpp} we give a specific example.
\end{remark}

In this section, our goal is to extend the theory of Section \ref{labsec:continuousobservation} but instead of the logarithmic price process $H$ that has continuous paths, we consider the process $\price$ that includes jumps. Again, as in Section \ref{labsec:discreteobservation}, we follow the same steps as in  Section \ref{labsec:continuousobservation}. Moreover, we will keep the same notation.\\

For $q \in \mathbb{Z}$ we keep the notation $\Gamma(q)$ for the process defined by  
\[
\Gamma_{-\pi}(q)=0, \quad
\Gamma_z(q)=\Gamma_z[d\price](q):= \frac{1}{2 \pi} \int_{-\pi}^{z}  e^{-\imath q t} d\price_t,\quad  z \in [-\pi,\pi].
\]

\begin{definition}\thlabel{labdef:FourierJumps}
Let 
\[
\fourier[d \price](q):= \Gamma_{\pi}(q).
\]
The system of coefficients $\{\fourierb_{N}[\vol]\}_{N \in \nn}$ is defined by 
\[
\fourierb_{N}[\vol](q) := 2 \pi \left\lbrace \bohrp{\fourier[d\price]}{\fourier[d\price]} \right\rbrace (q), \quad q \in \mathbb{Z}.
\]	
\end{definition}

\thref{labth:fundamentalobservation} takes the following form.
\begin{proposition}\thlabel{labth:fundamentalobservationJumps}
\begin{equation}\label{labeq:appbybohrconvoJumps}
\frac{1}{2\pi}\sum_{-\pi \leq t \leq \pi} e^{-\imath q t} \Delta J^2_t + \fourier[\vol](q) = \fourierb_{N}[\vol](q)-  \resi,
\end{equation}
where
\begin{equation}
\resi[t]:=\frac{2\pi}{2N +1}\sum_{|l| \leq N} \left\lbrace \int_{-\pi}^{t}\Gamma_{z-}(q-l) d\Gamma_z(l) + \int_{-\pi}^{t}\Gamma_{z-}(l) d\Gamma_z(q-l)\right\rbrace.
\end{equation}
\end{proposition}
\begin{remark}
Similar to how we did in Section \ref{labsec:continuousobservation}, we will prove that $\resi$ converges to zero. But now, equation \eqref{labeq:appbybohrconvoJumps} shows that $\fourierb_{N}[\vol](q)$ estimates the coefficient $\fourier[\vol](q)$  plus an additional term $\frac{1}{2\pi}\sum_{-\pi \leq t \leq \pi} e^{-\imath q t} \Delta J^2_t$. Hence, we do not recover the Fourier coefficients of the volatility $\vol=\sigma^2$. Yet, we will prove that under appropriate conditions, the trigonometric polynomial \eqref{labEq:FejerexpressionContinuous}, properly  rescaled, allows us to pathwise recover the process of quadratic jumps $\Delta J^2$; see \thref{labthm:PathUCjumps} and \thref{labcor:PathUCjumps}; see also  \thref{labthm:FFjumpfunction} in  Appendix \ref{labsecFFjumpfunction}.   
\end{remark}
\begin{remark}
Continuing with the above remark, we outline a possible approach to also obtain an indirect estimation   of $\vol$ from $\fourierb_{N}[\vol]$.   Indeed, the trigonometric polynomial $\frac{2 \pi}{M} \trigb_{N,M}[\vol]$ approximates  $\Delta J^2$.  Hence, the Fourier coefficients of  $\frac{2 \pi}{M} \trigb_{N,M}[\vol]$ approximate those of  $\Delta J^2$, and  discounting  from $\fourierb_{N}[\vol]$ one obtains a candidate to provide an approximation of $\fourier[\vol]$. We study this approach in future work.
\end{remark}
\subsection{Error's representation}
We define for $t \in [-\pi, \pi]$
\begin{align*}
\xi_s(q,N,t)&:=\int_{-\pi}^{s} e^{-\imath q {z}}  \tilde D_N({t}-{z}) d \price_{z}, \quad \xi_{-\pi}(q,N,t)=0,  \quad s \in [-\pi,t],\\
Y_t(q,N)&:= \frac{1}{2\pi} \int_{-\pi}^{t} \xi_{s-}(q,N,s)d\price_{s}, \quad Y_{-\pi}(q,N)=0,\\
Z_t(q,N)&:= \frac{1}{2\pi} \int_{-\pi}^{t} e^{-\imath q {s}} dY_{s}(0,N), \quad Z_{-\pi}(q,N)=0.
\end{align*}

\thref{labLemmaUpperbound} keeps its exact form:
\begin{align*}
2\pi \resi&= Y_{\pi}(q,N) +   Z_{\pi}(q,N).
\end{align*}

\subsection{A key estimation: The $\mathbb{L}_p$-norm of the remainder}\label{labsec:keyestimationJump}
In this section we fix $p > 2$ , $\alpha,\beta \in (1,\infty)$ with $\alpha^{-1} + \beta^{-1}=1$, and  $a_1,a_2,a_3>1$ such that $a_1^{-1}+a_2^{-1}+a_3^{-1}=1$. We use the convention that for an adapted process $X$, $X_{\infty}=0$.  Hence, for  $\{\tau_n\}_{n \in \nn}$  a sequence of stopping times taking values in $[-\pi,\pi]\cup\{\infty\}$ that exhaust the jump times of $J$ in $[-\pi,\pi]$ we then have that $X_{\tau_n} = X_{\tau_n} 1_{\{\tau_n<\infty\}} = X_{\tau_n} 1_{\{\tau_n \leq \pi\}}$. For a complex valued process $X=\Re X + \imath \Im X$ recall the norm $\triplenorm{X}$ in \eqref{labeq:triplenorm}. \\

\thref{labthmest1unbounded} takes the following form.

\begin{theorem}\thlabel{labthmest1unboundedJumps}
Let $\sigma$ satisfy \thref{lab:integrabilityforsigma} with exponent $\integrabilityratea=p(\alpha \vee \beta \vee a_2)$, and let $J$ satisfy \thref{labass:localpJumpsummability} with  exponents $\integrabilityratejumps>0$ and $\integrabilityratejumpsa= p(\beta \vee a_1 \vee a_2)$.  Furthermore, assume that the jump times $\{\tau_n\}_{n \in \nn}$ of $J$ satisfy
\begin{equation}\label{labeqest1unboundedJumps}
\sum_{n \in \nn} \sqrt[a_3]{E\left[  \Delta J_{\tau_n}^{2a_3}\right]}<\infty.
\end{equation}
Let
\begin{equation}
l := \min \left\lbrace  \frac{\integrabilityratejumps}{2+\integrabilityratejumps}, \frac{1}{2} \wedge \frac{\integrabilityratejumps}{4}, \frac{1}{2\beta}, \frac{\integrabilityratea-2}{3\integrabilityratea - 2}\right\rbrace.
\end{equation}
Then,
\begin{equation} \label{labal:upperboundresijump}
\triplenorm{\resi[]} =  O \left( \frac{1}{N^{l}} \right). 
\end{equation}
\end{theorem}
\begin{proof}
Similar to \thref{labthmest1unbounded}, in order to show \eqref{labal:upperboundresijump}, we only present how to obtain an upper bound for  $\left\| \sup_{-\pi \leq s \leq \pi} |\Re Y_s(q,n,N)| \right\|_{\mathbb{L}_p}$.
Thus, we focus on the estimation of $\triplenorm{\Re Y(q,N)}$.\\

For $t \in [-\pi,\pi]$ let 
\begin{align}
\aux^1_t:=
& \int_{-\pi}^{t} \Bigl(\int_{-\pi}^{s-} \cos(q z) \tilde D_N(z-s) \, dH_z \Bigr)   dH_s \label{labal:tt1}\\  
\aux^2_t:=
& \int_{-\pi}^{t} \Bigl(\int_{-\pi}^{s-} \cos(q z) \tilde D_N(z-s) \, dJ_z \Bigr)  dH_s  \label{labal:tt2} \\
\aux^3_t:=
&\int_{-\pi}^{t} \Bigl(\int_{-\pi}^{s-} \cos(q z) \tilde D_N(z-s) \, dH_z \Bigr)   dJ_s  \label{labal:tt3}\\ 
\aux^4_t:=
&\int_{-\pi}^{t} \Bigl(\int_{-\pi}^{s-} \cos(q z) \tilde D_N(z-s) \, dJ_z \Bigr)   dJ_s. \label{labal:tt4}
\end{align}

We have
\begin{align*}
(\Re Y (q,N))^*=   (\aux^1+\aux^2+\aux^3 + \aux^4)^*
\leq (\aux^1)^* +  (\aux^2)^* + (\aux^3)^* + (\aux^4)^*,
\end{align*}
where the supremum processes on the right-hand side are evaluated at $t=\pi$.  Hence
\begin{align*}
\left\| (\Re Y(q,N))^*_{t}\right\|_{L^p} \leq   \left\| (\aux^1)^*\right\|_{L^p} 
 + \left\| (\aux^2)^*\right\|_{L^p}  + \left\| (\aux^3)^*\right\|_{L^p}  + \left\| (\aux^4)^*\right\|_{L^p},
\end{align*}

We have $\left\| (\aux^1)^*\right\|_{L^p}=O \left( \frac{1}{N^{\frac{1}{2 \beta}}} \right)$ due to \thref{labthmest1unbounded}, in Section \ref{labsec:continuousobservation}.  Moreover,  $\left\| (\aux^2)^*\right\|_{L^p}=O \left( \frac{1}{N^{\frac{1}{2} \wedge \frac{\integrabilityratejumps}{4}}} \right)$, due to \thref{labproJumps:estimationT2}. The third term  $\aux^3$ satisfies $\left\| (\aux^3)^*\right\|_{L^p} = O \left( \frac{1}{N^{\frac{\integrabilityratea-2}{3\integrabilityratea - 2}}} \right)$, due to \thref{labproJumps:estimationT3}. Finally,  the last term  $\aux^4$  satisfies $\left\| (\aux^4)^*\right\|_{L^p}= O \left( \frac{1}{N^{\frac{\integrabilityratejumps}{2+\integrabilityratejumps}}} \right)$, due to \thref{labproJumps:estimationT4}.  
\end{proof}

In the following proposition we obtain an upper bound for  $\left\| (\aux^2)^*\right\|_{L^p}$ under the conditions of \thref{labass:localpJumpsummability} for the jumps and \thref{lab:integrabilityforsigma} for the diffusion coefficient.
\begin{proposition}\thlabel{labproJumps:estimationT2}
Let $\aux^2$ be defined by \eqref{labal:tt2}.  Assume the conditions of \thref{labthmest1unboundedJumps}. Then
\[
 \left\| (\aux^2)^*\right\|_{L^p} = O \left( \frac{1}{N^{\frac{1}{2} \wedge \frac{\integrabilityratejumps}{4}}} \right).
\]
\end{proposition}
\begin{proof}
Let $X_s:=\int_{-\pi}^{s} \cos(q z) \tilde D_N(z-s) \, dJ_z$.  We have
\[
\left\| (\aux^2)^*\right\|_{L^p} \leq \sqrt[p]{C_{p}}\sqrt[p]{E\left[ \left|  \int_{-\pi}^{\pi} X^2_{s-}   d[H]_s\right|^{\frac{p}{2}}\right]},
\]
due to BDG-inequality.  Moreover 
\[
\int_{-\pi}^{\pi} X_{s-}^2  \sigma_s^2 ds 
\leq
\sqrt[\alpha]{\int_{-\pi}^{\pi} |\sigma_s|^{2\alpha} ds } \sqrt[\beta]{\int_{-\pi}^{\pi} |X_{s-}|^{2\beta}  ds },
\]
due to H\"older inequality on $[-\pi, \pi]$ and then
\begin{align*}
E\left[ \left| \int_{-\pi}^{\pi}X_{s-}^2  \sigma_s^2 ds \right|^{\frac{p}{2}}\right] 
\leq
&E \left[ \left|  \int_{-\pi}^{\pi} |\sigma_s|^{2\alpha} ds \right|^{\frac{p}{2 \alpha}}  \left| \int_{-\pi}^{\pi} \left|X_{s-}\right|^{2\beta}  ds\right|^{\frac{p}{2 \beta}}\right] \\
\leq 
&\sqrt[\alpha]{E \left[  \left| \int_{-\pi}^{\pi} |\sigma_s|^{2\alpha} ds \right|^{\frac{p}{2}}\right] } 
\sqrt[\beta]{E \left[  \left| \int_{-\pi}^{\pi} \left|X_{s-}\right|^{2\beta} ds \right|^{\frac{p}{2}}\right] } \\
\leq
&\sqrt[\alpha]{E \left[  \left| \int_{-\pi}^{\pi} |\sigma_s|^{2\alpha} ds \right|^{\frac{p}{2}}\right] } 
(2\pi)^{\frac{p-2}{2\beta}} \sqrt[\beta]{E \left[  \int_{-\pi}^{\pi} \left| X_{s-}\right|^{p\beta} ds \right] }\\
=
&\sqrt[\alpha]{E \left[  \left| \int_{-\pi}^{\pi} |\sigma_s|^{2\alpha} ds \right|^{\frac{p}{2}}\right] } 
(2\pi)^{\frac{p-2}{2\beta}} \sqrt[\beta]{\int_{-\pi}^{\pi} E \left[\left| X_{s-}\right|^{p\beta}\right] ds  }.
\end{align*}
Hence, the proof concludes taking $\kappa=p \beta$ in \thref{lablem:auxestiJ}.
\end{proof}

Now we obtain an upper bound for $\left\| (\aux^3)^*\right\|_{L^p}$ with $\aux^3$  defined in \eqref{labal:tt3}. 
\begin{proposition}\thlabel{labproJumps:estimationT3}
Assume the conditions of \thref{labthmest1unboundedJumps}. Then
\[
\left\| (\aux^3)^*\right\|_{L^p} = O \left( \frac{1}{N^{\frac{\integrabilityratea-2}{3\integrabilityratea - 2}}} \right).
\]
\end{proposition}
\begin{proof}
Let $X_s:=\int_{-\pi}^{s} \cos(q z) \tilde D_N(z-s) \, dH_z$.  We have
\[
\left\| (\aux^3)^*\right\|_{L^p} \leq \sqrt[p]{C_{p}}\sqrt[p]{E\left[ \left|  \int_{-\pi}^{\pi} X^2_{s-}   d[J]_s\right|^{\frac{p}{2}}\right]},
\]
due to BDG-inequality.  Moreover
\[
\left|\int_{-\pi}^{\pi } X_{s-}^2 \,  d[J]_s \right|^{\frac{p}{2}}
\leq [J]_{\pi}^{\,\frac{p}{2}-1} \int_{-\pi}^{\pi} \left| X_{s-}\right|^p \, d[J]_s.
\]

Let $\{\tau_n\}$ exhaust the jump times of  $J$. Taking expectation, and interchanging summation with expectation:
\[
E\left[ \left|\int_{-\pi}^{\pi } X_{s-}^2 \,  d[J]_s \right|^{\frac{p}{2}}\right] 
\leq \sum_{n \in \nn} E\left[[J]_{\pi}^{\,\frac{p}{2}-1}|X_{\tau_n}|^p \Delta J_{\tau_n}^2\right].
\]
Let $a_1,a_2,a_3>1$ be such that $a_1^{-1}+a_2^{-1}+a_3^{-1}=1$. Then
\[
E\left[ \left(\int_{-\pi}^{\pi } X_{s-}^2 \,  d[J]_s \right)^{\frac{p}{2}}\right] 
\leq 
\sqrt[a_1]{E\left[  [J]_{\pi}^{\frac{p-2}{2}a_1}\right]}  \sum_{n \in \nn} \sqrt[a_2]{E\left[ |X_{\tau_n}|^{p a_2}\right]} \sqrt[a_3]{E\left[  \Delta J_{\tau_n}^{2a_3}\right]}.
\]

Now we estimate the term $E\left[ |X_{\tau_n}|^{a}\right]$ with $a=p a_2$.
For $\delta>0$ we use the notation: $\rho(\delta):=\max\{\tau_n -\delta,-\pi\}$. To start we have 
\begin{align*}
&E\left[ |X_{\tau_n}|^{a}\right]\\
\leq
&C_{a} E\left[ \left|  \int_{-\pi}^{\tau_n} \cos^2(q z) \tilde D_N^2(z - \tau_n) \sigma^2_z dz \right|^{\frac{a}{2}} \right]\\
&\leq
C_{a} 2^{a-1} E\left[ \left|  \int_{-\pi}^{\rho(\delta)} \cos^2(q z) \tilde D_N^2(z - \tau_n)  \sigma^2_zdz \right|^{\frac{a}{2}} \right]
+C_{a} 2^{a-1} E\left[ \left|  \int_{\rho(\delta)}^{\tau_n} \cos^2(q z) \tilde D_N^2(z - \tau_n) \sigma^2_zdz  \right|^{\frac{a}{2}} \right],
\end{align*}
for a suitable positive constant $C_{a}$. The first summand  $E\left[ \left|  \int_{-\pi}^{\rho(\delta)} \cos^2(q z) \tilde D_N^2(z - \tau_n)  \sigma^2_zdz \right|^{\frac{a}{2}} \right]$ can be estimated as follows. The rescaled Dirichlet kernel $\tilde D_N$ satisfies $\tilde D_N^2(x) \leq \frac{\pi^2}{\delta^2 N^2}$ for $|x| \in [-\pi,\pi]$ and $|x| \geq \delta$. Hence
\begin{align*}
E\left[ \left|  \int_{-\pi}^{\rho(\delta)} \cos^2(q z) \tilde D_N^2(z - \tau_n)  \sigma^2_zdz \right|^{\frac{a}{2}} \right]
&\leq 
\left( \frac{\pi^2}{\delta^2 N^2}\right)^{\frac{a}{2}}  E[[H]_{\pi}^{\frac{a}{2}}].
\end{align*}
The second summand satisfies
\begin{align*}
E\left[ \left|  \int_{\rho(\delta)}^{\tau_n} \cos^2(q z) \tilde D_N^2(z - \tau_n) \sigma^2_zdz  \right|^{\frac{a}{2}} \right]
&\leq 
E\left[ \left|  \int_{\rho(\delta)}^{\tau_n} d[H]_z \right|^{\frac{a}{2}} \right].
\end{align*}
Hence for $r, r'>1$ conjugate numbers
\begin{align*}
E\left[ \left|  \int_{\rho(\delta)}^{\tau_n} \cos^2(q z) \tilde D_N^2(z - \tau_n) \sigma^2_zdz  \right|^{\frac{a}{2}} \right]
&\leq 
E\left[ \left|  \int_{\rho(\delta)}^{\tau_n} d[H]_z \right|^{\frac{a}{2}} \right]\\
& \leq
\delta^{\frac{a}{2 r'}} E\left[ \left|  \int_{-\pi}^{\pi} \sigma^{2r}_z dz \right|^{\frac{a}{2r}} \right].
\end{align*}

For $\delta=N^{-b}$ we obtain an upper bound of the form 
\[
E\left[ |X_{\tau_n}|^{a}\right] = O\left( \frac{1}{N^{(1-b)a}}\right)  +  O\left( \frac{1}{N^{ \frac{a b}{2 r'}}}\right).
\]
For $b=\frac{2 r'}{2 r' +1}$ we obtain
\[
E\left[ |X_{\tau_n }|^{a}\right] 
=
O\left( \frac{1}{N^{ \frac{a}{2 r' +1}}}\right)
\]
Hence
\[
\sqrt[a_2]{E\left[ |X_{\tau_n}|^{a}\right]} =
O\left(\frac{1}{N^{\frac{p}{2 r' + 1}}}\right)=
O\left(\frac{1}{N^{p\frac{r-1}{3 r - 1}}}\right).
\]
\end{proof}

In the following proposition we obtain an upper bound  for $\left\| (\aux^4)^*\right\|_{L^p}$ with $\aux^4$  defined in \eqref{labal:tt4}. 
\begin{proposition}\thlabel{labproJumps:estimationT4}
Assume the conditions of \thref{labthmest1unboundedJumps}. Then
\[
\left\| (\aux^4)^*\right\|_{L^p} = 
O \left( \frac{1}{N^{\frac{\integrabilityratejumps}{2+\integrabilityratejumps}}} \right).
\]
\end{proposition} 
\begin{proof}
Let $X_s := \int_{-\pi}^{s} \cos(q z) \tilde D_N(z-s) \, dJ_z$.  First, we apply Jensen's inequality:
\[
\left(\int_{-\pi}^{\pi } X_{s-}^2 \,  d[J]_s \right)^{\frac{p}{2}}
\leq [J]_{\pi}^{\,\frac{p}{2}-1} \int_{-\pi}^{\pi} X_{s-}^p \, d[J]_s.
\]
Let $\{\tau_n\}$ exhaust the jump times  of $J$. Taking expectation, and interchanging summation with expectation:
\[
E\left[ \left|\int_{-\pi}^{\pi } X_{s-}^2 \,  d[J]_s \right|^{\frac{p}{2}}\right] 
\leq \sum_{n \in \nn} E\left[[J]_{\pi}^{\,\frac{p}{2}-1}|X_{\tau_n -}|^p \Delta J_{\tau_n }^2\right].
\]
Now take $a_1,a_2,a_3>1$ such that $a_1^{-1}+a_2^{-1}+a_3^{-1}=1$ then
\[
E\left[ \left(\int_{-\pi}^{\pi } X_{s-}^2 \,  d[J]_s \right)^{\frac{p}{2}}\right] 
\leq 
\sqrt[a_1]{E\left[ [J]_{ \pi}^{\frac{p-2}{2}a_1}\right]}  
\sum_{n \in \nn} \sqrt[a_2]{E\left[ |X_{\tau_n -}|^{p a_2}\right]} \sqrt[a_3]{E\left[  \Delta J_{\tau_n}^{2a_3}\right]}.
\]

Now we estimate the term $E\left[ |X_{\tau_n -}|^{a}\right]$ with $a=p a_2$. Arguments are similar to that of \thref{labproJumps:estimationT3}, for completeness we give the details.
For $\delta>0$ we use the notation: $\rho(\delta):=\max\{\tau_n -\delta,-\pi\}$. To start we have 
\begin{align*}
&E\left[ |X_{\tau_n -}|^{a}\right]\\
\leq
&C_{a} E\left[ \left|  \int_{-\pi}^{\tau_n -} \cos^2(q z) \tilde D_N^2(z - \tau_n)  d[J]_z \right|^{\frac{a}{2}} \right]\\
\leq
&C_{a} 2^{a-1} E\left[ \left|  \int_{-\pi}^{\rho(\delta)} \cos^2(q z) \tilde D_N^2(z - \tau_n)  d[J]_z \right|^{\frac{a}{2}} \right]
+C_{a} 2^{a-1} E\left[ \left|  \int_{\rho(\delta)+}^{\tau_n -} \cos^2(q z) \tilde D_N^2(z - \tau_n)  d[J]_z \right|^{\frac{a}{2}} \right],
\end{align*}
for a suitable positive constant $C_{a}$. The first summand  $E\left[ \left|  \int_{-\pi}^{\rho(\delta)} \cos^2(q z) \tilde D_N^2(z - \tau_n)  d[J]_z \right|^{\frac{a}{2}} \right]$ can be estimated as follows. The rescaled Dirichlet kernel $\tilde D_N$ satisfies $\tilde D_N^2(x) \leq \frac{\pi^2}{\delta^2 N^2}$ for $|x| \in [-\pi,\pi]$ and $|x| \geq \delta$. Hence
\begin{align*}
E\left[ \left|  \int_{-\pi}^{\rho(\delta)} \cos^2(q z) \tilde D_N^2(z - \tau_n)  d[J]_z \right|^{\frac{a}{2}} \right] 
&\leq 
\left( \frac{\pi^2}{\delta^2 N^2}\right)^{\frac{a}{2}}  E[[J]_{\pi}^{\frac{a}{2}}].
\end{align*}
The second summand satisfies
\begin{align*}
E\left[ \left|  \int_{\rho(\delta)+}^{\tau_n -} \cos^2(q z) \tilde D_N^2(z - \tau_n)  d[J]_z \right|^{\frac{a}{2}} \right]
&\leq 
E\left[ \left|  \int_{\rho(\delta)+}^{\tau_n -} d[J]_z \right|^{\frac{a}{2}} \right].
\end{align*}
Hence
\begin{align*}
\left\|  \int_{\rho(\delta)+}^{\tau_n -} \cos^2(q z) \tilde D_N^2(z - \tau_n)  d[J]_z \right\|_{L_{a/2}}
&=O(\delta^{\integrabilityratejumps}),
\end{align*}
where the estimate follows from  \eqref{labasseq2:localpJumpsummabilitylefttaun} since $\frac{a}{2} \leq \integrabilityratejumpsa$. For $\delta=N^{-b}$ we obtain an upper bound of the form 
\[
E\left[ |X_{\tau_n -}|^{a}\right] = O\left( \frac{1}{N^{(1-b)a}}\right)  +  O\left( \frac{1}{N^{ \frac{1}{2} a b \integrabilityratejumps}}\right).
\]
For $b=\frac{2}{2+\integrabilityratejumps}$ we obtain
\[
E\left[ |X_{\tau_n -}|^{a}\right] 
=
O\left(\frac{1}{N^{a \frac{ \integrabilityratejumps}{2+\integrabilityratejumps}}}\right).
\]
Hence
\[
\sqrt[a_2]{E\left[ |X_{\tau_n -}|^{a}\right]} =
O\left(\frac{1}{N^{p \frac{ \integrabilityratejumps}{2+\integrabilityratejumps}}}\right).
\]
\end{proof}

\subsection{Coefficients' convergence} \label{labsec:coefficientconvergencejumps}
Following the reasoning of Section \ref{labsec:coefficientconvergence}, once we have the estimation in  \thref{labthmest1unboundedJumps} we deduce the convergence of coefficients. We state the theorem and omit the proof since the details are similar.\\

\thref{labthm:coeffConvergence} takes the following form.
\begin{theorem}\thlabel{labthm:coeffConvergencejumps}
Let  $l$ be as in \thref{labthmest1unboundedJumps}. 
Take $\pow \in (0,l)$,  $p >  2$ and $r >0$. If $\pow p-r>1$  and the conditions of \thref{labthmest1unboundedJumps} hold true, then for a constant $\tilde K>0$ and for $M=O(N^r)$
\[
P\left(\left\lbrace \sup_{|q| \leq M}  \left| \fourier[\vol] (q)+\frac{1}{2\pi}\sum_{-\pi \leq t \leq \pi} e^{-\imath q t} \Delta J^2_t-\fourierb_{N}[\vol](q)\right| 
\geq
6 {\tilde K} N^{\pow -l} \right\rbrace \right) = O\left( \frac{1}{N^{\pow p -r}}\right).
\]
Hence, in the complement of a null event 
\begin{align*}
\sup_{|q| \leq M} \left\lbrace \left| \fourier [\vol](q)+\frac{1}{2\pi}\sum_{-\pi \leq t \leq \pi} e^{-\imath q t} \Delta J^2_t-\fourierb_{N}[\vol](q)\right| \right\rbrace   < 6 {\tilde K} N^{\pow -l}, \quad \text{eventually}.
\end{align*}
In particular
\[
\lim_{N\to \infty} \fourierb_{N}[\vol](q)= \fourier[\vol](q)+\frac{1}{2\pi}\sum_{-\pi \leq t \leq \pi} e^{-\imath q t} \Delta J^2_t, \quad a.s.
\]
\end{theorem}

\subsection{Uniform convergence of trigonometric polynomials}\label{labsec:polynomialconvergencejumps}
Let $\trig_M[\Delta J^2]$ be the trigonometric polynomial determined by the system of `exact coefficients' of the quadratic jump process $\Delta J^2$ , that is
\begin{align}
\trig_{M}[\Delta J^2](t)&=\sum_{|l|\leq M}\left(1-\frac{|l|}{M} \right) \fourierj[\Delta J ^2](l) e^{\imath l t} \label{labeq:trigvolexaccoeffjumps}\\
\fourierj[\Delta J^2](l)&= \frac{1}{2 \pi}\sum_{z \in [-\pi, \pi]} e^{-\imath l z} \Delta J_z^2.
\end{align}
Moreover, let $\trigb_{N,M}$ be the trigonometric polynomial  constructed with the coefficients $\fourierb$ in \thref{labdef:FourierJumps}.

The next theorem is the analogous result to \thref{labthm:PathUC}. It establishes that, under an appropriate growth rate for $M$, the difference between the trigonometric polynomials $\frac{1}{M}\trigb_{N,M}$ and $\frac{1}{M}\trig_{M}[\Delta J^2]$ converges to zero uniformly.
\begin{theorem}\thlabel{labthm:PathUCjumps}
Let $l$ be as in  \thref{labthm:coeffConvergencejumps}.
Take $\pow \in (0,l)$,  $p >  2$ and $r >0$. If $r< (l - \pow )\wedge (\pow p-1)$  and the conditions of \thref{labthmest1unboundedJumps} hold true, then, for $M=O(N^r)$	
\begin{equation}\label{labeq:uniformconvergenceofapproxPolyjumps}
\lim_{\substack{N,M\to \infty \\ M=O(N^r)}}\sup_{t \in [-\pi,\pi]} \frac{1}{M}	\left|\trig_{M}[\Delta J^2](t) - \trigb_{N,M}(t)\right| = 0, a.s.
\end{equation}
\end{theorem}
\begin{proof}
Assume that $M \leq K N^r$. For $t \in [-\pi,\pi]$ 
\begin{align*}
\left|\trig_{M}[\Delta J^2](t) - \trigb_{N,M}(t)\right|
&=
\left|\sum_{|q|\leq M}\left(1-\frac{|q|}{M} \right) \left\lbrace  \fourierj[\Delta J^2](q) - \fourierb_N(q) \right\rbrace  e^{\imath q t}\right|\\
& \leq 
\left|\sum_{|q|\leq M}\left(1-\frac{|q|}{M} \right) \left\lbrace  \fourierj[\Delta J^2](q) + \fourier[\vol](q) - \fourierb_N(q) \right\rbrace  e^{\imath q t}\right|\\
&+ \left|\sum_{|q|\leq M}\left(1-\frac{|q|}{M} \right) \fourier[\vol](q)  e^{\imath q t}\right|\\
& \leq
M \sup_{|q| \leq M}   \left| \fourierj[\Delta J^2](q) + \fourier[\vol](q)- \fourierb_N[\vol](q) \right| + B,
\end{align*}
where  $B$ is a positive random variable that do not depend on time and $B<\infty$ a.s. Hence,
\begin{align*}
\frac{1}{M}\left|\trig_{M}[\Delta J^2](t) - \trigb_{N,M}(t)\right|
&\leq \sup_{|q| \leq M}   \left| \fourierj[\Delta J^2](q) + \fourier[\vol](q)- \fourierb_N[\vol](q) \right| + \frac{B}{M} \\
& \leq  6 {\tilde K} N^{\pow -l} + \frac{B}{M}, \textit{ eventually},
\end{align*}
where  the second inequality holds true eventually in the complement of a null event due to \thref{labthm:coeffConvergencejumps}.
\end{proof}

\begin{corollary}\thlabel{labcor:PathUCjumps}
Under the conditions of \thref{labthm:PathUCjumps}, for almost all $\omega$, the trigonometric polynomial $\frac{2 \pi}{M}\trigb_{N,M}$ converges pointwise to $\Delta J^2$.
\end{corollary}
\begin{proof}
This is a consequence of \thref{labthm:PathUCjumps} above, and of \thref{labthm:FFjumpfunction} in Appendix \ref{labsecFFjumpfunction}.
\end{proof}
\subsection{Case study: Compensated Poisson process}\label{labsubs:cpp}
Let $\poissonp$ be a Poisson process with intensity $\tilde \lambda>0$ constructed as follows. Let $T_1 , T_2 , \ldots$ be a sequence of independent, exponentially distributed random variables with parameter $\tilde \lambda> 0$.  Let $\tau_0=0$ and $\tau_n:=T_1+\ldots+ T_n$.  Define the Poisson process $\poissonp$ by $\poissonp_t:= \max\{n \geq 0 \mid \tau_n \leq t\}$.  Let  $\{Y_n\}_{n \in \nn}$ be an i.i.d. sequence of random variables, independent of $\poissonp$,  with $\left\| Y_1\right\|_{\infty} = 1$.  Let $\lambda=\tilde \lambda E[Y_1]$, and  $J$ be the purely discontinuous local martingale defined by 
\begin{equation}\label{labeqcpp}
J_t=\sum_{n=1}^{\infty} Y_n 1_{[\tau_n,\infty)}(t) - \lambda t.
\end{equation}
Now we translate the definition of $J$ so that it is indexed by the interval $[-\pi,\pi]$.\\

We verify that $\{J_t\}_{-\pi \leq t \leq \pi}$ satisfies the conditions of \thref{labass:localpJumpsummability} and of \thref{labthmest1unboundedJumps}.  We start with \eqref{labasseq:localpJumpsummability} in \thref{labass:localpJumpsummability}. 
\begin{proposition}\thlabel{labpro:poissonjumpsA}
The process $J$ defined in \eqref{labeqcpp} satisfies \eqref{labasseq:localpJumpsummability}. More precisely, for $\integrabilityratejumpsa \in \nn\setminus \{1\}$, and $0<\delta <1$,
\begin{equation*}
E \left[ \left( M_t(\delta) \right)^{\integrabilityratejumpsa}  \right]=
O\left( \delta \right).
\end{equation*}
\end{proposition}
\begin{proof}
For $\delta \in(0,1)$ let $Z$ be a random variable with distribution  $Poisson(r)$ for a parameter $r \leq 2 \tilde \lambda \delta$. Then for any $p\in \nn$ with $p>1$ we have 
\begin{align*}
E[(M_t(\delta))^p]  
= 
E\left[\left( \sum_{\substack{z \in [-\pi, \pi]\setminus\{t\}\\ 0 < |t-z| < \delta}}\Delta J_z^2\right)^p  \right]
\leq 
E\left[\left( \sum_{\substack{z \in [-\pi, \pi]\setminus\{t\}\\ 0 < |t-z| < \delta}}\Delta \poissonp_z^2\right)^p  \right]
\leq &E[Z^p].
\end{align*}
Hence 
\[
E[(M_t(\delta))^p]   \leq \sum_{k=1}^{p} \left\{ \begin{matrix} p \\ k \end{matrix} \right\} (2 \tilde \lambda \delta)^k \leq  \delta \sum_{k=1}^{p} \left\{ \begin{matrix} p\\ k \end{matrix} \right\}(2 \tilde \lambda)^k,
\]
where the coefficients in the sums are the Stirling numbers of the second kind.
\end{proof}

We continue with \eqref{labasseq2:localpJumpsummabilitylefttaun}.  Let $\{\tau_n\}_{n \in \nn}$ be the sequence of stopping times introduced in the definition of $J$. 
\begin{proposition}
For $\integrabilityratejumpsa \in \nn\setminus \{1\}$, and $0<\delta <1$, we have
\begin{equation}
E\left[ \left| \int_{(\tau_n -\delta,\tau_n)\cap [-\pi, \pi]} d[J]_z \right|^{\integrabilityratejumpsa}\right]  =O(\delta).
\end{equation}
\end{proposition}
\begin{proof}
The number of jumps in the interval $(\tau_n-\delta,\tau_n)\cap[-\pi,\pi]$  is estimated from above by  $L_n(\delta):=\#\{k < n \mid \tau_{n}- \tau_{k} < \delta\}$.  Taking $\tau_n$ as initial point and reversing time, we see that $L_n$ is estimated from above by a Poisson random variable with an intensity proportional to $\delta$.  Now we conclude as in \thref{labpro:poissonjumpsA}.
\end{proof}

We conclude verifying \eqref{labeqest1unboundedJumps} in  \thref{labthmest1unboundedJumps}. From this property, condition \eqref{labasseq3:localpJumpsummability} in \thref{labass:localpJumpsummability} will follow.
\begin{proposition}
For any $\kappa>1$
\[
\sum_{n \in \nn} \sqrt[\kappa]{E\left[  |\Delta J_{\tau_n} 1_{\{\tau_n \leq \pi\}}|^{2 \kappa} \right]}<\infty.
\]
\end{proposition}
\begin{proof}
For any $\pow>1 \vee \kappa $ we  have $P(\poissonp_{\pi} \geq n)=O(n^{-\pow})$. Hence, there is a constant $K_{\pow}>0$ such that 
\begin{align*}
\sqrt[\kappa]{E\left[  |\Delta J_{\tau_n}|^{2\kappa} 1_{\{\tau_n \leq \pi\}}\right]}
&\leq \sqrt[\kappa]{P({\{\tau_n \leq \pi\}})}\\
&\leq 
K_{\pow} n^{-\frac{\pow}{\kappa}}.
\end{align*}
Now the result follows from a simple comparison.
\end{proof}

\subsection{Numerical illustrations: Poisson process}\label{labsubs:cppnumericalillust}
In this section we  present a numerical exercise\footnote{All the results were obtained with the code in \href{https://github.com/ErTrAg/Fourier-Malliavin-Mancino-estimator-of-volatility-with-jumps}{https://github.com/ErTrAg/Fourier-Malliavin-Mancino-estimator-of-volatility-with-jumps}} illustrating the approximation result in \thref{labcor:PathUCjumps}.\\

We simulate a logarithmic price process $\price$ and estimate the rescaled trigonometric polynomial $\frac{2 \pi}{M}\trigb_{N,M}$. This numerical exercise will clearly illustrate the pointwise  convergence  to $\Delta J^2$. For concreteness we take the process 
\[
\price_t = \int_{-\pi}^{t}\sigma(\sin(s)+2)dW_s + \poissonp_t - \lambda t,\quad \lambda=2, \sigma=1,
\]
where  $\poissonp$ is a Poisson process with compensator rate $\lambda$. \\

\begin{figure}[H]
	\centering
	\begin{subfigure}{0.38\linewidth}
		\centering
		\includegraphics[width=\linewidth]{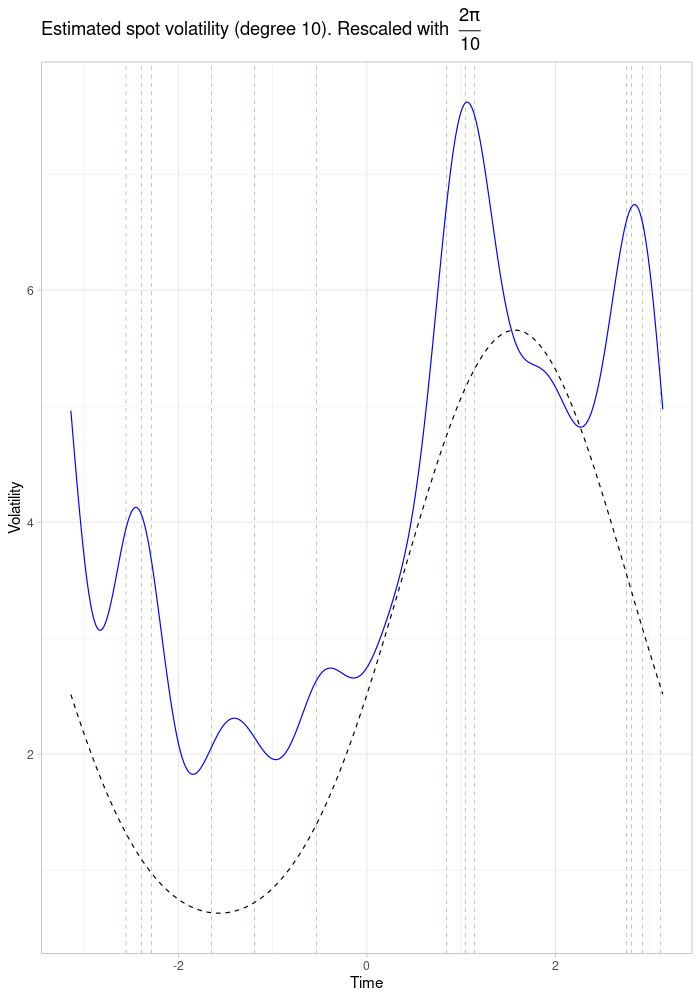}
		\caption{The Fourier estimator of spot volatility with 10 degrees.}
		\label{labf3}
	\end{subfigure}
	\hfill
	\begin{subfigure}{0.38\linewidth}
		\centering
		\includegraphics[width=\linewidth]{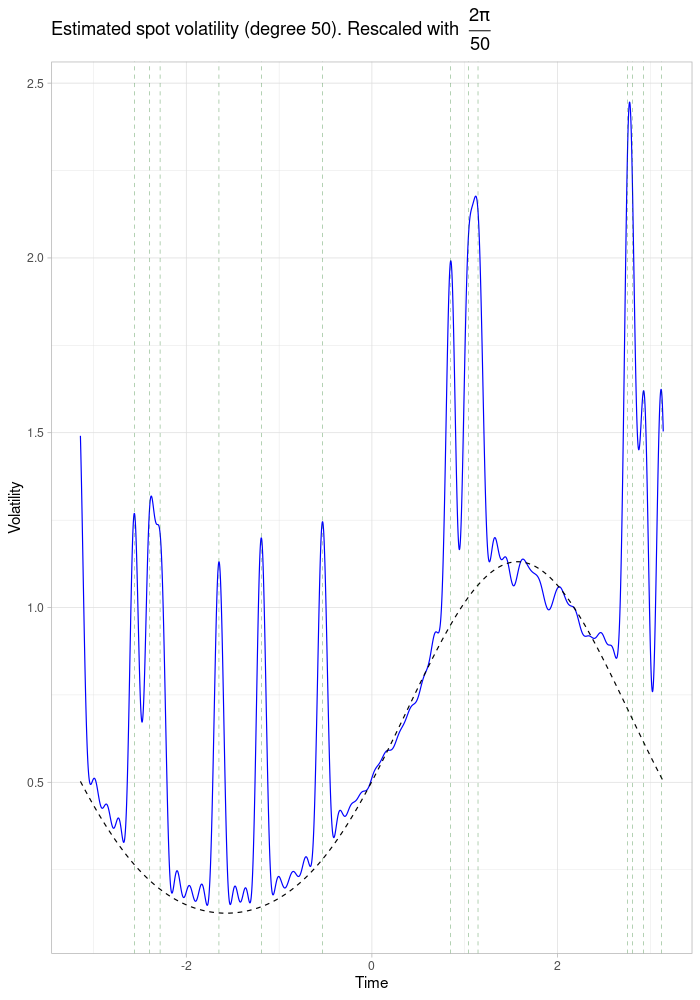}
		\caption{The Fourier estimator of spot volatility with 50 degrees.}
		\label{labf4}
	\end{subfigure}
	\caption{The rescaled trigonometric polynomial $\frac{2 \pi}{M}\trigb_{N,M}$, degrees 10 and 50, respectively.}
\end{figure}

\begin{figure}[H]
	\centering
	\begin{subfigure}{0.38\linewidth}
		\centering
		\includegraphics[width=\linewidth]{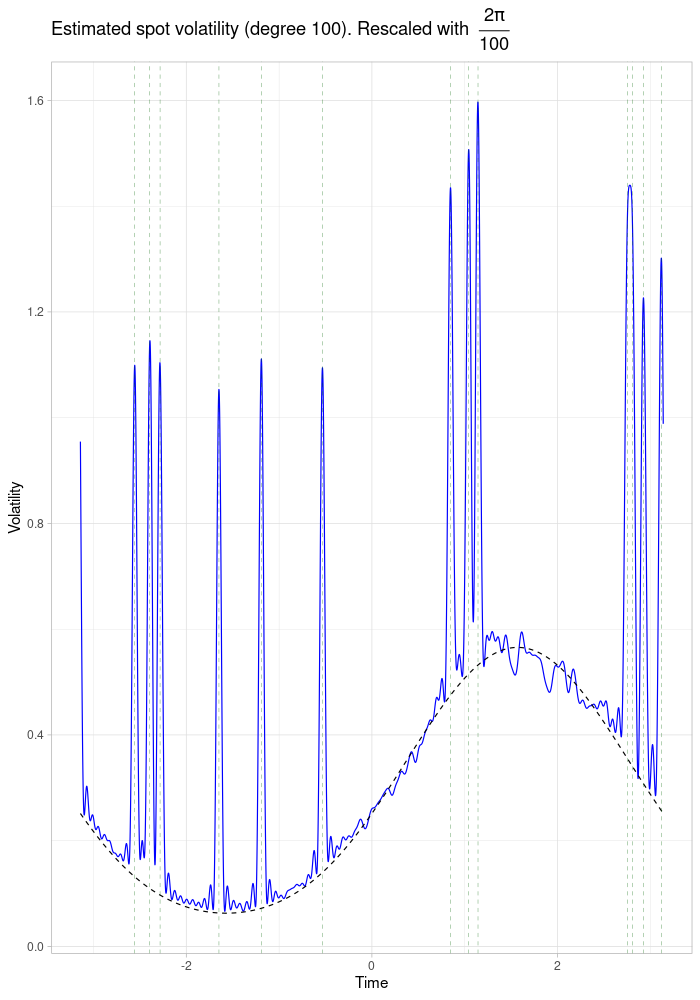}
		\caption{The Fourier estimator of spot volatility with 100 degrees.}
		\label{labf5}
	\end{subfigure}
	\hfill
	\begin{subfigure}{0.38\linewidth}
		\centering
		\includegraphics[width=\linewidth]{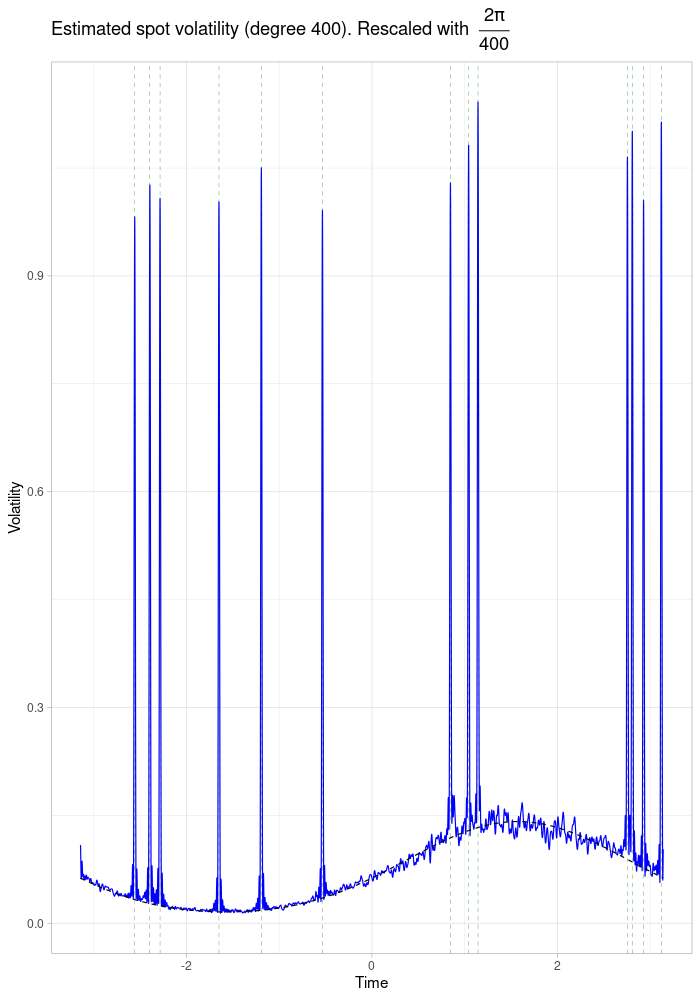}
		\caption{The Fourier estimator of spot volatility with 400 degrees.}
		\label{labf6}
	\end{subfigure}
	\caption{The rescaled trigonometric polynomial $\frac{2 \pi}{M}\trigb_{N,M}$, degrees 100 and 400, respectively.}
\end{figure}

In Figures \ref{labf3} to \ref{labf6},  we see in blue line the rescaled trigonometric polynomial $\frac{2 \pi}{M}\trigb_{N,M}$,
the Fourier estimator of spot volatility, with degrees 10, 50, 100 and 400, respectively. Note that these figures visually illustrate what we have proved about  the pointwise convergence of the polynomials to the jump process $\Delta J^2$,  which in this case coincides with $\Delta \poissonp$.

\appendix
\section{Few elementary facts on Fourier coefficients}
\subsection{The Dirichlet kernel}
Recall that the \textbf{Dirichlet kernel} was introduced in \eqref{labEq:DirichletKernel}. It is given by $D_N(t):= \sum_{|l| \leq N} \exp (\imath l t)$. It is well known that
\[
D_N(t)=\frac{\sin(Nt+t/2)}{\sin(t/2)}.
\]
Recall  the \textbf{rescaled Dirichlet kernel} in equation \eqref{labEq:RescaledDirichletKernel}. It is given by $\tilde D_N(t):=\frac{1}{2N +1} D_N(t)$.
In the proof of \thref{labthmest1unbounded}  we applied Lemma  \ref{labLemma:DirichletkernelEstimation} below. The statement is just  Exercise 3.1.6 of \cite{Grafakos2014}.
\begin{lemma}\thlabel{labLemma:DirichletkernelEstimation}
	Take $r>1$. For the Dirichlet kernel $D_N$ there exist positive constants $b_r$ and $B_r$ such that for every $N \in \nn$, the following inequality holds true:
	\[
	b^r_r(2N+1)^{r-1} \leq  \int_{-\pi}^{\pi}\left|D_N(s)\right|^r ds \leq B^r_r(2N+1)^{r-1}.
	\]
\end{lemma}
\subsection{The Fej\'er kernel}
Recall that the Fej\'er kernel is defined by 
\[
\fejer_N(t):=\frac{1}{N}\sum_{j=0}^{N-1} D_N(t).
\]
It satisfies 
\[
\fejer_N(t)=\frac{1}{N}\left( \frac{\sin(N t/2)}{\sin(t/2)} \right)^2. 
\]
From this expression one easily obtains that for $\delta>0$ the following useful inequality holds true
\[
\sup_{\substack{z\in[-\pi,\pi],\\ |z|\geq \delta}} \fejer_N(z) \leq \frac{\pi^2}{\delta^2 N}.
\]
We will use this inequality in the proof of \thref{labthm:FFjumpfunction} below.
\subsection{The Dirichlet kernel in discrete time}
In this section we give a discretized version of \thref{labLemma:DirichletkernelEstimation}. For completeness we give full details in the proof. For $r>1$ recall the constant
$\dddot A_r:= 5 + \frac{2\pi^r}{r-1}$ defined in \eqref{labeq:Aconstant}.

\begin{lemma}\thlabel{labLemma:DirichletkernelEstimationdiscretized}
Take a partition $\nu=\{-\pi=s_0<s_1<\ldots<s_l = \pi\}$ of the interval $[-\pi,\pi]$ and let $\rho$ be its norm: $\rho=\max_{i=0,\ldots,l-1}{|s_{i+1} - s_{i}|}$. Define  $\cutabs{t}:=\max\{z \in \nu \mid z \leq t\}$.  Take $r>1$. For  $t \in [-\pi,\pi]$  we have the upper bound
\[
\int_{-\pi}^{t}\left| D_N(\cutabs{t}-\cutabs{s})\right|^r  ds 
\leq 
5(\rho + (2N+1)^{-1})(2N+1)^r + 2 \frac{\pi^{r}}{r-1} (2N+1)^{r-1}.
\]
In particular
\begin{align*}
\int_{-\pi}^{t}\left|\tilde D_N(\cutabs{t}-\cutabs{s})\right|^r  ds 
&\leq 
5(\rho + (2N+1)^{-1}) + 2 \frac{\pi^{r}}{r-1} (2N+1)^{-1}\\
&=5\rho + (2N+1)^{-1} \left(\frac{2\pi^{r}}{r-1} + 5\right)\\
&\leq 5\rho + (2N+1)^{-1}\dddot A_r.
\end{align*}
\end{lemma}
\begin{proof}
For $s \in [-\pi,\pi]$ we define $\rig{s}:=\inf\{x \in \nu \mid x \geq s\}$ with the convention $\inf \emptyset = -\infty$. The Dirichlet kernel admits on $[-\pi,\pi]$ the upper bound $|D_N(\cdot)| \leq (2N+1) \wedge f(\cdot)$, where
\[
f(s):=\begin{cases}
\frac{\pi}{|s|} & \quad 0<|s| \leq  \pi\\
\infty & \quad	s=0.
\end{cases}
\]
Take $t \in [-\pi,\pi]$ and denote $t_0:=\cutabs{t}$, $\delta:=(2N+1)^{-1}$. We present the proof only for the case $t_0>0$, since the case $t_0 \leq 0$ is easier.
\begin{enumerate}
\item Let $t_1:= \cutabs{t_0-\pi-\delta}$. Only the case $t_1 \geq -\pi$ is non-trivial in this part.  We estimate the integral $I_1:=\int_{-\pi}^{t_1}\left| D_N(\cutabs{t}-\cutabs{s})\right|^r  ds$.  For $s \in [-\pi,t_1]$ we have  $\pi + \delta \leq  t_0 -\cutabs{s}$. Hence $0\leq 2 \pi -(t_0 -\cutabs{s})\leq  \pi -\delta $ and 
\[
\left| D_N(t_0 -\cutabs{s})\right|=
\left| D_N(2\pi - t_0+ \cutabs{s})\right|
\leq (2N+1) \wedge \frac{\pi}{\left|2\pi - t_0 +  \cutabs{s}\right|}.
\]
Thus, the integral $I_1$  is bounded from above by  $\int_{-\pi}^{t_1} (2N+1)^r \wedge \frac{\pi^r}{\left|2\pi - t_0 + \cutabs{s}\right|^r}ds$ and this last integral can be estimated as follows.

Let $t_m \in \nu$ be such that  $t_m=\rig{\delta -\pi + \rho}$.  Consider the case $t_m \leq t_1$; otherwise the next estimations simplify. We  have
\begin{align*}
\int_{-\pi}^{t_1} (2N+1)^r \wedge \frac{\pi^r}{(2\pi - t_0 + \cutabs{s} )^r}  ds 
&\leq 
(t_{m}+\pi)(2N+1)^r + \int_{t_m}^{t_1} \frac{\pi^r}{(2\pi - t_0 + {s} -\rho)^r}  ds\\
&\leq 
(\delta + \rho)(2N+1)^r + \int_{t_m}^{t_1} \frac{\pi^r}{(2\pi - t_0 + {s} -\rho)^r}  ds, 
\end{align*}
where in the first inequality we used the fact that for $s \in [t_m,t_1]$ 
\[
2\pi - t_0 + \cutabs{s}\geq 2\pi - t_0 + {s} -\rho \geq 2\pi - t_0 + t_m -\rho \geq \delta.
\]
Moreover
\begin{align*}
\int_{t_m}^{t_1}  \frac{\pi^r}{(2\pi - t_0 + {s} -\rho)^r}  ds
&= 
\frac{\pi^r}{1-r}\left[ (2\pi - t_0 + {s} -\rho)^{1-r}\right]_{t_m}^{t_1}\\
&\leq  \frac{\pi^r}{r-1}(2\pi - t_0 + t_m -\rho)^{1-r}\\
&\leq  \frac{\pi^r}{r-1}\delta^{1-r}.
\end{align*}
Thus,
\begin{align*}
I_1
&\leq  (\delta + \rho)(2N+1)^r + \frac{\pi^r}{r-1}\delta^{1-r}.
\end{align*}
\item Let $t_2 :=\rig{t_0-\pi+\delta}$.  We estimate the integral $I_2:=\int_{t_2}^{t}\left| D_N( t_0 - \cutabs{s})\right|^r  ds$. For $s \in [t_2,t]$ note that $\cutabs{s} \geq t_2$ and then we have  $0 \leq t_0-\cutabs{s} \leq \pi -\delta$. Let $t_k \in \nu$ be such that  $t_k=\rig{t_0- \delta}$. Note that for $s \leq t_{k-1}$
\[
\delta < t_0 - t_{k-1} \leq t_0 -s \leq t_0 -\cutabs{s}.
\] 
Only the case $t_{k-1} \geq t_2$ is interesting for the next estimations. We have
\begin{align*}
I_2 &\leq \int_{t_2}^{t} (2N+1)^r \wedge \frac{\pi^r}{(t_0 - \cutabs{s} )^r}ds\\
&\leq 
\int_{t_2}^{t_{k-1}}\frac{\pi^r}{(t_0 - \cutabs{s} )^r}ds + \int_{t_{k-1}}^{t}(2N+1)^r ds\\
&\leq 
\int_{t_2}^{t_{k-1}}\frac{\pi^r}{(t_0 - {s} )^r}ds + (t-t_{k-1})(2N+1)^r\\
&=\frac{\pi^r}{r-1}\left[ (t_0 - {s} )^{1-r} \right]_{t_2}^{t_{k-1}} + (t-t_{k-1})(2N+1)^r\\
&\leq 
\frac{\pi^r}{r-1}(t_0-t_{k-1})^{1-r} + (2\rho + \delta)(2N+1)^r\\
&\leq 
\frac{\pi^r}{r-1}\delta^{1-r} + 2(\rho + \delta)(2N+1)^r.
\end{align*}
\item Now we estimate the integral $I_3=\int_{t_1}^{t_2} \left| D_N(t_0-\cutabs{s})\right|^r  ds$ in the case $t_2 > t_1$. We have
\begin{align*}
t_2-t_1
& = t_2-(t_0-\pi+\delta) + t_0-\pi+\delta -\cutabs{t_0-\pi-\delta}\\
&\leq \rho + t_0-\pi-\delta  -\cutabs{t_0-\pi-\delta} +2\delta  \\
&\leq 2 \rho + 2 \delta.
\end{align*}
Now 
\begin{align*}
\int_{t_1}^{t_2} \left| D_N(\cutabs{t}-\cutabs{s})\right|^r  ds
&\leq 2 (\delta + \rho)(2N+1)^r.
\end{align*}
\item As a consequence 
\begin{align*}
\int_{-\pi}^{t}\left| D_N(\cutabs{t}-\cutabs{s})\right|^r  ds 
&\leq I_1+I_2+I_3 \leq 5 (2N+1)^r (\delta  + \rho) + 2 \frac{\pi^r}{r-1}\delta^{1-r}.
\end{align*}
\end{enumerate}
\end{proof}
\section{Fourier-Fej\'er inversion of a jump function}\label{labsecFFjumpfunction}
Let $\mho:[-\pi,\pi] \to \rn$ be a cadlag function with  jump function $\Delta \mho: [-\pi,\pi]\to \rn$ defined as usual by $\Delta \mho_t:=\mho_{t}-\mho_{t-}$. Assume that the sum of quadratic  jumps is finite: $[\mho]:=\sum_{z \in [-\pi, \pi]} \Delta \mho_z^2 < \infty$. 
In this section we develop a ``Fourier-Fejér inversion theory'' for $\Delta \mho$. To this end, we start with the analogous definitions in this context. Let
\begin{align*}
\fourierj[\Delta \mho^2](q)&:= \frac{1}{2\pi}\sum_{z \in [-\pi, \pi]} e^{-\imath q z} \Delta \mho_z^2\\
S_N[\Delta \mho^2](t)&:=\sum_{|q| \leq N} \fourierj[\Delta \mho^2](q) e^{\imath q t}, and,\\
\trig_N[\Delta \mho^2]&:=\frac{1}{N}\sum_{j=0}^{N-1} S_j[\Delta \mho^2]. 
\end{align*}
For a function $f$, the ``convolution'' with the Fej\'er kernel is
\[
f \star  \fejer_N(t) := \frac{1}{2\pi}\sum_{z \in [-\pi, \pi]} f(z) \fejer_N(t-z).
\]
Through this convolution, the trigonometric polynomial $\trig_N[\Delta \mho^2]$ can be expressed as
\[
\trig_N[\Delta \mho^2](t) = \Delta \mho^2 \star  \fejer_N(t).
\]

Now that we have the preliminary notation ready, we establish the  main result of this section.
\begin{theorem}\thlabel{labthm:FFjumpfunction}
Define for $\delta \in (0,\pi)$ and $t \in [-\pi, \pi]$
\[
M_t(\delta):= \sum_{\substack{z \in [-\pi, \pi] \setminus \{t\} \\ 0< |z-t|< \delta}}\Delta \mho_z^2.
\]
Take $t \in [-\pi,\pi]$ and assume that 
\[
\lim_{\delta \to 0 } M_t(\delta) =0.
\]
Then, 
\begin{align*}
\lim_{N \to \infty} \left| \frac{2 \pi }{N} \trig_N[\Delta \mho^2](t) - \Delta \mho^2_t\right| 
=0.
\end{align*}
Moreover, for $O \subset [-\pi,\pi]$ an open set with 
\[
\lim_{\delta \to 0 }  \sup_{t \in O} M_t(\delta) = 0,
\]
we have the uniform convergence
\begin{align*}
\lim_{N \to \infty} \sup_{t \in O}\left| \frac{2 \pi }{N} \trig_N[\Delta \mho^2](t) - \Delta \mho^2_t\right| 
&=0.
\end{align*}
\end{theorem}
\begin{proof}
Take $t \in [-\pi,\pi]$.   We consider only the case where there is a jump at time $t$: $\Delta \mho^2_t>0$, the case $\Delta \mho^2_t=0$ is simpler.  Note that 
\[
2\pi \trig_N[\Delta \mho^2](t)= \Delta \mho^2_t \fejer_N(0)
+ \sum_{\substack{z \in [-\pi,\pi]\setminus \{t\}  \\ 0<|z-t|<\delta}} \Delta \mho^2_z \fejer_N(t-z)
+ \sum_{\substack{z \in [-\pi,\pi]\setminus \{t\}  \\ |z-t|\geq \delta}} \Delta \mho^2_z \fejer_N(t-z).
\]
We have
\[
\sum_{\substack{z \in [-\pi,\pi]\setminus \{t\}  \\ 0<|z-t|<\delta}} \Delta \mho^2_z \fejer_N(t-z) \leq N M_t(\delta).
\]
Recall that for $\delta>0$ we have $\sup_{z\in[-\pi,\pi] \setminus \{t\}, |z-t|\geq \delta} \fejer_N(t-z) \leq \frac{\pi^2}{\delta^2 N}$. Hence
\[
\sum_{\substack{z \in [-\pi,\pi]\setminus \{t\}, \\ |z-t|\geq \delta}} \Delta \mho^2_z \fejer_N(t-z) \leq [\mho] \sup_{\substack{z\in [-\pi,\pi]\setminus \{t\}, \\ |z-t|\geq \delta}} \fejer_N(t-z) \leq [\mho] \frac{\pi^2}{\delta^2 N}.
\]
As a consequence we obtain:
\begin{align*}
\left| \frac{2 \pi }{N} \trig_N[\Delta \mho^2](t) - \Delta \mho^2_t\right| 
&\leq 
M_t(\delta) + [\mho] \frac{\pi^2}{\delta^2 N^2}.
\end{align*}
In particular, taking $\delta=N^{-\frac{1}{2}}$ we obtain
\begin{align*}
\left| \frac{2 \pi}{N} \trig_N[\Delta \mho^2](t) - \Delta \mho^2_t\right| 
&\leq 
M_t(N^{-\frac{1}{2}}) + [\mho] \frac{\pi^2}{N}.
\end{align*}
This shows the pointwise convergence. For the uniform convergence we simply take the supremum over $t \in O$ on both sides of the inequality and conclude with the assumption on $\sup_{t \in O}M_t$.
\end{proof}

\section{Auxiliary estimations}
\begin{lemma}\thlabel{lablem:auxestiH}
Take $\kappa  > 2$ and $\alpha,\beta \in (1,\infty)$ with $\frac{1}{\alpha} +\frac{1}{\beta}=1$. Let $\sigma$ satisfy \thref{lab:integrabilityforsigma} with exponent $\integrabilityratea =\kappa  \vee 2 \alpha$. Take $z \in [-\pi,\pi]$. Let $X_t(z):=\int_{-\pi}^{t} \cos(q s) \tilde D_N(z-s) \, dH_s$, for $t \in [-\pi,z]$,  We have
\begin{equation}\label{labal:estimationforkdHOnotation}
E \left[ \left|X_{t}(z)\right|^{\kappa}\right]	\leq C_{\kappa} B_{2\beta}^{\kappa} E\left[\left( \int_{-\pi}^{\pi} |\sigma_s|^{2 \alpha} ds \right)^{\frac{\kappa}{2 \alpha}} \right] \frac{1}{N^{\frac{\kappa}{2 \beta}}}, \text{ for } -\pi \leq t \leq z.
\end{equation}
\end{lemma}
\begin{proof}
We have 
\begin{align*}
E\left[ \sup_{-\pi \leq t \leq z}\left|  X_{t}(z)\right|^{\kappa}\right]
\leq  C_{\kappa} E\left[\left\langle X(z)\right\rangle_{z}^{\frac{\kappa}{2}} \right],
\end{align*}
due to the BDG-Inequality. Moreover
\begin{align*}
E\left[\left\langle X(z)\right\rangle_{t}^{\frac{\kappa}{2}} \right]
\leq   
& E\left[\left(\sqrt[\alpha]{\int_{-\pi}^{z} |\sigma_s|^{2 \alpha} ds} \sqrt[\beta]{\int_{-\pi}^{z}|\tilde D_N(z-s)|^{2 \beta} ds}  \right)^{\frac{\kappa}{2}} \right]\\
\leq  
&E\left[\left( \int_{-\pi}^{\pi} |\sigma_s|^{2 \alpha} ds \right)^{\frac{\kappa}{2 \alpha}} \right] \left( \int_{-\pi}^{\pi}|\tilde D_N(z-s)|^{2 \beta} ds\right)^{\frac{\kappa}{2 \beta}}\\
\leq  
&E\left[\left( \int_{-\pi}^{\pi} |\sigma_s|^{2 \alpha} ds \right)^{\frac{\kappa}{2 \alpha}} \right] \left( \int_{-\pi}^{\pi}|\tilde D_N(s)|^{2 \beta} ds\right)^{\frac{\kappa}{2 \beta}}, 
\end{align*}
the first inequality holds true due to Holder's inequality. The second is clear since $\tilde D_N$ is deterministic. The last inequality is clear.

The integral $E\left[\left( \int_{-\pi}^{\pi} |\sigma_s|^{2 \alpha} ds \right)^{\frac{\kappa}{2 \alpha}} \right]$ is finite due to \thref{lab:integrabilityforsigma} with exponent $\kappa \vee 2\alpha$.
Moreover
\begin{align*}
\int_{-\pi}^{\pi}|\tilde D_N(s)|^{2 \beta} ds 
&\leq 
B_{2\beta}^{2\beta} (2N+1)^{-1},
\end{align*}
due to \thref{labLemma:DirichletkernelEstimation}. Hence
\begin{align*}
E\left[\sup_{-\pi \leq t \leq z}\left|  X_{t}(z)\right|^{\kappa}\right]
\leq  
C_{\kappa} E\left[\left(\int_{-\pi}^{\pi} |\sigma_s|^{2 \alpha} ds \right)^{\frac{\kappa}{2 \alpha}} \right] 
B_{2\beta}^{\kappa} (2N+1)^{-\frac{\kappa}{2 \beta}}.
\end{align*}
\end{proof}

\begin{lemma}\thlabel{lablem:auxestiJ}
Take $\kappa>2$ and let $J$ be a process that satisfies \thref{labass:localpJumpsummability} with  $\integrabilityratejumpsa \geq\frac{\kappa}{2}$ and $\integrabilityratejumps>0$. Let $X_s:=\int_{-\pi}^{s} \cos(q z) \tilde D_N(z-s) \, dJ_z$.  We have for a fixed $C>0$
\begin{equation}\label{labal:estimationforkdJ}
E \left[ \left|X_{s-}\right|^{\kappa}\right]
\leq  
C_{\kappa} 2^{\frac{\kappa}{2}-1} \left( {\pi}{N^{-\frac{1}{2}}}\right)^{\kappa}  E \left[ \left(\sum_{z \in [-\pi, \pi]}  \Delta J^2_z \right)^{\frac{\kappa}{2} }  \right] + C C_{\kappa} 2^{\frac{\kappa}{2}-1} \left( N^{-\frac{\integrabilityratejumps}{4}}\right)^{\kappa}. 
\end{equation}
More simple notation:
\begin{equation}\label{labal:estimationforkdJOnotation}
E \left[ \left|X_{s-}\right|^{\kappa}\right]
= O\left( \frac{1}{N^{\frac{\kappa}{2}}}\right) + O\left( \frac{1}{N^{\frac{\kappa \integrabilityratejumps}{4}}}\right).
\end{equation}
\end{lemma}
\begin{proof}
We have for some $C_{\kappa}>0$ by the Burkholder-Davis-Gundy inequality  that
\begin{align*}
E \left[\left|X_{s-}\right|^{\kappa}\right] 
&\leq 
C_{\kappa} E \left[ \left(\int_{-\pi}^{s-} \tilde D_N^2(z-s) \, d[J]_z\right)^{\frac{\kappa}{2} }\right].
\end{align*}
We have for $\delta>0$
\begin{align*}
&E \left[ \left(\int_{-\pi}^{s-} \tilde D_N^2(z-s) \, d[J]_z\right)^{\frac{\kappa}{2} }\right]\\
\leq
2^{\frac{\kappa}{2}-1} &E \left[ \left(\sum_{\substack{z \in [-\pi, \pi\wedge s) \\  0< |z-s|< \delta}}  \tilde D_N^2(z-s) \Delta J^2_z \right)^{\frac{\kappa}{2} } + \left(\sum_{ \substack{z \in [-\pi, \pi\wedge s) \\|z-s|\geq  \delta}} \tilde D_N^2(z-s) \Delta J^2_z \right)^{\frac{\kappa}{2} } \right],
\end{align*}
due to the elementary inequality $(a+b)^r \leq 2^{r - 1}(|a|^{r}+|b|^{r})$, for $r=\frac{\kappa}{2}>1$.\\

The  ``distant jumps'' $E \left[ \left(\sum_{\substack{z \in [-\pi, \pi \wedge s) \\ |z-s|\geq \delta}}  \tilde D_N^2(z-s) \Delta J^2_z \right)^{\frac{\kappa}{2} }  \right]$ are estimated as follows. For $\delta>0$ we have $\sup_{z\in[-\pi,\pi \wedge s), |z-s|\geq \delta} D_N(z-s) \leq \frac{\pi}{\delta}$. Hence, for the rescaled Dirichlet kernel $\tilde D_N$ we have  $\tilde D_N^2(x) \leq \frac{\pi^2}{\delta^2 N^2}$ for $|x| \geq \delta$. Thus,
\begin{align*}
E \left[ \left(\sum_{\substack{z \in [-\pi, \pi \wedge s ) \\ |z-s| \geq \delta}} \tilde D_N^2(z-s) \Delta J^2_z \right)^{\frac{\kappa}{2} }  \right]
&\leq 
\left( \frac{\pi}{\delta N}\right)^{\kappa}  E \left[ \left(\sum_{\substack{z \in [-\pi, \pi \wedge s ) \\ |z-s| \geq \delta}}  \Delta J^2_z \right)^{\frac{\kappa}{2} }  \right].
\end{align*}

We continue with the ``near located jumps''  $E \left[ \left(\sum_{z \in [-\pi, \pi \wedge s), 0<|z-s|< \delta} \tilde D_N^2(z-s) \Delta J^2_z \right)^{\frac{\kappa}{2}} \right]$. We have
\begin{align*}
E \left[ \left(\sum_{\substack{z \in [-\pi, \pi \wedge s) \\ 0<|z-s|< \delta}} \tilde D_N^2(z-s) \Delta J^2_z \right)^{\frac{\kappa}{2} }  \right]
&\leq 
E \left[ \left(\sum_{\substack{z \in [-\pi, \pi \wedge s) \\ 0< |z-s|< \delta}} \Delta J^2_z \right)^{\frac{\kappa}{2} }  \right].
\end{align*}

Then, for a constant $C>0$
\begin{align*}
E \left[ \left(\sum_{\substack{z \in [-\pi, \pi \wedge s) \\ 0<|z-s|< \delta}} \tilde D_N^2(z-s) \Delta J^2_z \right)^{\frac{\kappa}{2} }  \right]
&\leq 
C \left(\delta^{\integrabilityratejumps}\right)^{\frac{\kappa}{2}},
\end{align*}
since we assume that \eqref{labasseq:localpJumpsummability} in \thref{labass:localpJumpsummability} holds true with $\integrabilityratejumpsa \geq \frac{\kappa}{2}$. As a consequence
\begin{align*}
E \left[ \left|X_{s-}\right|^{\kappa}\right]
&\leq  C_{\kappa} 2^{\frac{\kappa}{2}-1} \left( \frac{\pi}{\delta N}\right)^{\kappa}  
E \left[ \left(\sum_{z \in [-\pi, \pi]}  \Delta J^2_z \right)^{\frac{\kappa}{2} }  \right] + C_{\kappa} 2^{\frac{\kappa}{2}-1} C\left( \delta^{\integrabilityratejumps}\right)^{\frac{\kappa}{2}}. 
\end{align*}
In particular,  for $\delta=N^{-\frac{1}{2}}$ the inequalities \eqref{labal:estimationforkdJ} and \eqref{labal:estimationforkdJOnotation} are direct consequence.
\end{proof}
\section{Chebyshev's inequality for higher moments}
The following is a form of Chebyshev's inequality for higher moments and it follows from Markov's inequality. 
\begin{proposition}\thlabel{labpro:Chebyshev}
Let $X$ be an integrable random variable. For $p>0$ and $t>0$:
\[
P\left( |X-E[X]| > t \sqrt[p]{E\left[ |X-E[X]|^p\right] }\right)  \leq \frac{1}{t^p}.
\]
\end{proposition}
\section{Diffusions}
\thref{labthmest1unbounded} requires the integrability condition in \thref{lab:integrabilityforsigma}. The  quality of an approximation through trigonometric polynomials can be quantitatively described with the modulus of continuity as it was discussed in \thref{labrem:moduluscontinuity}.  In this section, we present a large class of diffusions  that provide many examples satisfying these conditions.\\

The next result follows from \cite[Theorems 2.3 and 2.4]{Gihman72} as a particular case.
\begin{theorem}\thlabel{labtheasydiffusions}
Let $a, b: [-\pi,\pi] \times \rn \to \rn$ be continuous functions. Assume they are locally Lipschitz continuous: there exists $L>0$
\[
|a(t,x)-a(t,y)| +  |b(t,x)-b(t,y)|  \leq L|x-y|,
\]
for $|x|, |y| \leq N$. They have at most quadratic growth
\[
|a(t,x)|^2 + |b(t,x)|^2 \leq K^2(1+|x|^2).
\]
Then, there exists a unique strong solution to the stochastic differential equation
\[
dX=a(t,X_t)dt + b(t,X_t)dW_t, X_{-\pi}=x \in \rn.
\]
Moreover, there exists a constant $C>0$ depending only on $m, K$ and $2\pi$ for which
\[
E[X^{2m}_{t}] \leq E[1+ x^{2m}]e^{Ct}, t \in [-\pi,\pi].
\]
\end{theorem}

\begin{corollary}\thlabel{labtheasydiffusionscorollaryintegrability}
Under the conditions of \thref{labtheasydiffusions}, we have for $r>1$ 
\[
E\left[ \int_{-\pi}^{\pi}\left| b(z,X_z)\right|^{r}  dz\right]< \infty.
\]
\end{corollary}
\begin{proof}
We have the upper bound:
\begin{align*}
\left( |b(t,X_t)|^2\right)^r 
&\leq K^{2r} 2^{r-1}\left(1+|X_t|^{2r}\right),
\end{align*}
by assumption on the coefficient $b$. Hence,
\begin{align*}
E\left[ \int_{-\pi}^{\pi} |b(t,X_t)|^{2r} dt \right] 
&\leq K^{2r} 2^{r-1}\left(2 \pi + E\left[ \int_{-\pi}^{\pi} |X_t|^{2r} dt\right]  \right)\\
&= K^{2r} 2^{r-1}\left(2 \pi + \int_{-\pi}^{\pi} E\left[ |X_t|^{2r}\right]  dt \right)\\
& < \infty,
\end{align*}
where the equality holds by the Tonelli-Fubini theorem and the second inequality due to \thref{labtheasydiffusions}. 
\end{proof}
\begin{corollary}\thlabel{labtheasydiffusionscorollarymodulusofcontinuity}
In addition to the conditions of \thref{labtheasydiffusions}, assume there is a Lipschitz constant  $L>0$ with 
\[
|b(t,x)-b(s,y)|  \leq L|x-y| + L|t-s|.
\]
Then, the process $\sigma_t:=b(t,X_t)$ has locally $\gamma$-H\"older continuous paths with $\gamma \in (0,\frac{1}{2})$.
\end{corollary}
\begin{proof}
For $\alpha>2$ and $-\pi<s<t<\pi$
\begin{align*}
E\left[ \left| \sigma_t - \sigma_s\right|^{\alpha}\right] 
&\leq K E\left[ \left|X_t-X_s\right|^{\alpha} +  \left| t - s \right|^{\alpha}\right]  \\
&= 
K  E\left[ \left| \int_{s}^{t} \sigma_z dW_z\right|^{\alpha}\right] + 
K \left| t - s \right|^{\alpha},
\end{align*}
for a suitable constant $K>0$.
Hence, for $p,q>1$ with $\frac{1}{p} + \frac{1}{q}=1$
\begin{align*}
E\left[ \left| \sigma_t - \sigma_s\right|^{\alpha}\right] 
&\leq 
K C_{\alpha} E\left[\left| \int_{s}^{t} |\sigma_z|^2 dz\right|^{\frac{\alpha}{2}}\right] + K \left| t - s \right|^{\alpha}\\
&\leq 
K C_{\alpha} \left| t-s\right|^{\frac{\alpha}{2 p}}  E\left[\left| \int_{-\pi}^{\pi} |\sigma_z|^{2 q} dz \right|^{\frac{\alpha}{2 q}}\right] +  K \left| t - s \right|^{\alpha}.
\end{align*}
As a consequence, the process $\{b(t,X_t)\}_{t\in [-\pi,\pi]}$ has a modification that is locally $\gamma$-H\"older continuous due to the Kolmogorov-{\v C}entsov continuity theorem; see e.g., \cite[2.2.8]{Karatzas1991}. The modification is actually indistinguishable since the process already has continuous paths. 
\end{proof}
	
\section*{Conflict of interest}
All authors declare no conflicts of interest in this paper.

\end{document}